\documentclass[aps,superscriptaddress]{revtex4-2}
\usepackage{amsmath}
\usepackage{units,xspace}
\usepackage{graphicx}
\usepackage{array}
\usepackage{slashed}
\usepackage{url,hyperref}
\usepackage{multirow}
\usepackage[utf8]{inputenc}
\usepackage[margin=1.25in]{geometry}
\usepackage[usenames,dvipsnames]{color}

\usepackage{datetime}

\newcommand{\gd}{\ensuremath{g_{\textrm D}}\xspace}

\DeclareGraphicsExtensions{.pdf,.png,.jpg,.jpeg}
\graphicspath{{./}{./fig/}}

\def\beq{\begin{equation}}
\def\eeq#1{\label{#1}\end{equation}}
\def\eeqn{\end{equation}}

\newenvironment{Eqnarray}%
   {\arraycolsep 0.14em\begin{eqnarray}}{\end{eqnarray}}
\def\beqa{\begin{Eqnarray}}
\def\eeqa#1{\label{#1}\end{Eqnarray}}
\def\eeqan{\end{Eqnarray}}

\let\bar=\overbar

\def\lsim{\mathrel{\raise.3ex\hbox{$<$\kern-.75em\lower1ex\hbox{$\sim$}}}}
\def\gsim{\mathrel{\raise.3ex\hbox{$>$\kern-.75em\lower1ex\hbox{$\sim$}}}}

\def\del{\partial}
\def\Dslash{\not{\hbox{\kern-4pt $D$}}}
\def\dslash{\not{\hbox{\kern-2pt $\del$}}}
\def\pslash{\not{\hbox{\kern-2pt $p$}}}
\def\ETmiss{\not{\hbox{\kern-4pt $E$}}_T}

\def\Dlr{\mathrel{\raise1.5ex\hbox{$\leftrightarrow$\kern-1em\lower1.5ex\hbox{$D$}}}}

\def\MSB{{\bar{M \kern -2pt S}}}
\def\msb{{\bar{\scriptsize M \kern -1pt S}}}

\def\drb{{\bar{\scriptsize D \kern -1pt R}}}

\newcommand\snowmass{\begin{center}\rule[-0.2in]{\hsize}{0.01in}\\\rule{\hsize}{0.01in}\\
\vskip 0.1in Submitted to the  Proceedings of the US Community Study\\ 
on the Future of Particle Physics (Snowmass 2021)\\ 
\rule{\hsize}{0.01in}\\\rule[+0.2in]{\hsize}{0.01in} \end{center}}

\begin{document}


\sloppy




\title{\large \vspace{-1cm} MoEDAL-MAPP -- an LHC Dedicated Detector Search Facility}

\author{B.~Acharya}
\altaffiliation[Also at ]{\footnotesize Int. Centre for Theoretical Physics, Trieste, Italy}   
\affiliation{\footnotesize Theoretical Particle Physics \& Cosmology Group, Physics Dept., King's College London, UK}
\author{J.~Alexandre}
\affiliation{\footnotesize Theoretical Particle Physics \& Cosmology Group, Physics Dept., King's College London, UK}
\author{P.~Benes}
\affiliation{\footnotesize IEAP, Czech Technical University in Prague, Czech~Republic}
\author{B.~Bergmann}
\affiliation{\footnotesize IEAP, Czech Technical University in Prague, Czech~Republic}
\author{S.~Bertolucci}
\affiliation{\footnotesize INFN, Section of Bologna, Bologna, Italy}
\author{A.~Bevan}
\affiliation{\footnotesize School of Physics and Astronomy, Queen Mary University of London, UK}
\author{H.~Branzas}
\affiliation{\footnotesize Institute of Space Science, Bucharest - M\u{a}gurele, Romania}
\author{P.~Burian}
\affiliation{\footnotesize IEAP, Czech Technical University in Prague, Czech~Republic}
\author{M.~Campbell}
\affiliation{\footnotesize Experimental Physics Department, CERN, Geneva, Switzerland}
\author{S.~Cecchini}
\affiliation{\footnotesize INFN, Section of Bologna, Bologna, Italy}
\author{Y.~M.~Cho}
\affiliation{\footnotesize Center for Quantum Spacetime, Sogang University, Seoul, Korea} 
\author{M.~de~Montigny}
\affiliation{\footnotesize Physics Department, University of Alberta, Edmonton, Alberta, Canada}
\author{A.~De~Roeck}
\affiliation{\footnotesize Experimental Physics Department, CERN, Geneva, Switzerland}
\author{J.~R.~Ellis}
\altaffiliation[Also at ]{\footnotesize National Institute of Chemical Physics \& Biophysics, Tallinn, Estonia} 
\affiliation{\footnotesize Theoretical Particle Physics \& Cosmology Group, Physics Dept., King's College London, UK}
\affiliation{\footnotesize Theoretical Physics Department, CERN, Geneva, Switzerland}\
\author{M.~El~Sawy}
\altaffiliation[Also at ]{\footnotesize Dept. of Physics, Faculty of Science, Beni-Suef University, Egypt} 
\affiliation{\footnotesize Experimental Physics Department, CERN, Geneva, Switzerland}
\author{M.~Fairbairn}
\affiliation{\footnotesize Theoretical Particle Physics \& Cosmology Group, Physics Dept., King's College London, UK}
\author{D.~Felea}
\affiliation{\footnotesize Institute of Space Science, Bucharest - M\u{a}gurele, Romania}
\author{M.~Frank}
\affiliation{\footnotesize Department of Physics, Concordia University, Montr\'{e}al, Qu\'{e}bec,  Canada}
\author{J.~Hays}
\affiliation{\footnotesize School of Physics and Astronomy, Queen Mary University of London, UK}
\author{A.~M.~Hirt}
\affiliation{\footnotesize Department of Earth Sciences, Swiss Federal Institute of Technology, Zurich, Switzerland}
\author{P.~Q.~Hung}
\affiliation{\footnotesize Department of Physics, University of Virginia, Charlottesville, Virginia, USA}
\author{J.~Janecek}
\affiliation{\footnotesize IEAP, Czech Technical University in Prague, Czech~Republic}
\author{M.~Kalliokoski}
\affiliation{\footnotesize Helsinki Institute of Physics, University of Helsinki, Helsinki, Finland}
\author{A.~Korzenev}
\affiliation{\footnotesize D\'epartement de Physique Nucl\'eaire et Corpusculaire, Universit\'e de Gen\`eve, Geneva, Switzerland}
\author{D.~H.~Lacarr\`ere}
\affiliation{\footnotesize Experimental Physics Department, CERN, Geneva, Switzerland}
\author{C.~Leroy}
\affiliation{\footnotesize D\'{e}partement de Physique, Universit\'{e} de Montr\'{e}al, Qu\'{e}bec, Canada}
\author{G.~Levi} 
\affiliation{\footnotesize INFN, Section of Bologna \& Department of Physics \& Astronomy, University of Bologna, Italy}
\author{A.~Lionti}
\affiliation{\footnotesize D\'epartement de Physique Nucl\'eaire et Corpusculaire, Universit\'e de Gen\`eve, Geneva, Switzerland}
\author{A.~Margiotta}
\affiliation{\footnotesize INFN, Section of Bologna \& Department of Physics \& Astronomy, University of Bologna, Italy}
\author{R.~Mase\l ek}
\affiliation{\footnotesize Institute of Theoretical Physics, University of Warsaw, Warsaw, Poland}
\author{A.~Maulik}
\affiliation{\footnotesize INFN, Section of Bologna, Bologna, Italy}
\affiliation{\footnotesize Physics Department, University of Alberta, Edmonton, Alberta, Canada}
\author{N.~Mauri}
\affiliation{\footnotesize INFN, Section of Bologna \& Department of Physics \& Astronomy, University of Bologna, Italy}
\author{N.~E.~Mavromatos}
\altaffiliation[Also at ]{\footnotesize Department of Physics, School of Applied Mathematical and Physical Sciences, 
National Technical University of Athens, Athens, Greece} 
\affiliation{\footnotesize Theoretical Particle Physics \& Cosmology Group, Physics Dept., King's College London, UK}
\author{M.~Mieskolainen}
\affiliation{\footnotesize Physics Department, University of Helsinki, Helsinki, Finland}
\author{L.~Millward}
\affiliation{\footnotesize School of Physics and Astronomy, Queen Mary University of London, UK}
\author{V.~A.~Mitsou}
\affiliation{\footnotesize IFIC, Universitat de Val\`{e}ncia - CSIC, Valencia, Spain}
\author{E.~Musumeci}
\affiliation{\footnotesize IFIC, Universitat de Val\`{e}ncia - CSIC, Valencia, Spain}
\author{R.~Orava}
\affiliation{\footnotesize Physics Department, University of Helsinki, Helsinki, Finland}
\author{I.~Ostrovskiy}
\affiliation{\footnotesize Department of Physics and Astronomy, University of Alabama, Tuscaloosa, Alabama, USA}
\author{P.-P.~Ouimet}
\affiliation{\footnotesize Physics Department, University of Regina, Regina, Saskatchewan, Canada}
\author{J.~Papavassiliou}
\affiliation{\footnotesize IFIC, Universitat de Val\`{e}ncia - CSIC, Valencia, Spain}
\author{L.~Patrizii}
\affiliation{\footnotesize INFN, Section of Bologna, Bologna, Italy}
\author{G.~E.~P\u{a}v\u{a}la\c{s}}
\affiliation{\footnotesize Institute of Space Science, Bucharest - M\u{a}gurele, Romania}
\author{J.~L.~Pinfold}
\email[\textbf{Corresponding author: }]{jpinfold@ualberta.ca}
\affiliation{\footnotesize Physics Department, University of Alberta, Edmonton, Alberta, Canada}
\author{L.~A.~Popa}
\affiliation{\footnotesize Institute of Space Science, Bucharest - M\u{a}gurele, Romania}
\author{V.~Popa}
\affiliation{\footnotesize Institute of Space Science, Bucharest - M\u{a}gurele, Romania}
\author{M.~Pozzato}
\affiliation{\footnotesize INFN, Section of Bologna, Bologna, Italy}
\author{S.~Pospisil}
\affiliation{\footnotesize IEAP, Czech Technical University in Prague, Czech~Republic}
\author{A.~Rajantie}
\affiliation{\footnotesize Department of Physics, Imperial College London, UK}
\author{R.~Ruiz~de~Austri}
\affiliation{\footnotesize IFIC, Universitat de Val\`{e}ncia - CSIC, Valencia, Spain}
\author{Z.~Sahnoun}
\affiliation{\footnotesize INFN, Section of Bologna, Bologna, Italy}
\affiliation{\footnotesize Research Centre for Astronomy, Astrophysics and Geophysics, Algiers, Algeria}
\author{M.~Sakellariadou}
\affiliation{\footnotesize Theoretical Particle Physics \& Cosmology Group, Physics Dept., King's College London, UK}
\author{K.~Sakurai}
\affiliation{\footnotesize Institute of Theoretical Physics, University of Warsaw, Warsaw, Poland}
\author{A.~Santra}
\affiliation{\footnotesize IFIC, Universitat de Val\`{e}ncia - CSIC, Valencia, Spain}
\author{S.~Sarkar}
\affiliation{\footnotesize Theoretical Particle Physics \& Cosmology Group, Physics Dept., King's College London, UK}
\author{G.~Semenoff}
\affiliation{\footnotesize Department of Physics, University of British Columbia, Vancouver, British Columbia, Canada}
\author{A.~Shaa}
\affiliation{\footnotesize Physics Department, University of Alberta, Edmonton, Alberta, Canada}
\author{G.~Sirri}
\affiliation{\footnotesize INFN, Section of Bologna, Bologna, Italy}
\author{K.~Sliwa}
\affiliation{\footnotesize Department of Physics and Astronomy, Tufts University, Medford, Massachusetts, USA}
\author{R.~Soluk}
\affiliation{\footnotesize Physics Department, University of Alberta, Edmonton, Alberta, Canada}
\author{M.~Spurio}
\affiliation{\footnotesize INFN, Section of Bologna \& Department of Physics \& Astronomy, University of Bologna, Italy}
\author{M.~Staelens}
\affiliation{\footnotesize Physics Department, University of Alberta, Edmonton, Alberta, Canada}
\author{M.~Suk}
\affiliation{\footnotesize IEAP, Czech Technical University in Prague, Czech~Republic}
\author{M.~Tenti}
\affiliation{\footnotesize INFN, CNAF, Bologna, Italy}
\author{V.~Togo}
\affiliation{\footnotesize INFN, Section of Bologna, Bologna, Italy}
\author{J.~A.~Tuszy\'{n}ski}
\affiliation{\footnotesize Physics Department, University of Alberta, Edmonton, Alberta, Canada}
\author{A.~Upreti}
\affiliation{\footnotesize Department of Physics and Astronomy, University of Alabama, Tuscaloosa, Alabama, USA}
\author{V.~Vento}
\affiliation{\footnotesize IFIC, Universitat de Val\`{e}ncia - CSIC, Valencia, Spain}
\author{O.~Vives}
\affiliation{\footnotesize IFIC, Universitat de Val\`{e}ncia - CSIC, Valencia, Spain}

\collaboration{THE MoEDAL-MAPP  COLLABORATION}
\noaffiliation

\date{\today}

\begin{abstract}
\noindent
\begin{center}
ABSTRACT \\ 
\end{center}
During LHC's Run-2 the MoEDAL experiment, the LHC's first dedicated search experiment, took over 6~fb$^{-1}$ of  data at IP8,  with  \emph{p-p} and Pb-Pb collisions, operating with 100\% efficiency. The LHCC has endorsed  the LoI of the MoEDAL Collaboration describing an exciting program  to  expand the search  for highly ionizing particles in proton-proton and heavy-ion collisions to include feebly ionizing and long-lived messengers of new physics at   LHC's Run-3 and eventually High Luminosity LHC (HL-LHC). In 2021, as part of Phase-1 of the MoEDAL-MAPP project,   the baseline  MoEDAL detector was approved by CERN for  reinstallation for Run-3  with increased efficiency, lower threshold, a factor of five increase in instantaneous luminosity, and,  a higher centre-of-mass energy.  MoEDAL will  continue the search for highly ionizing particle avatars of new physics with an emphasis  on the  search for massive, single and multiply electrically charged particles arising from, for example, supersymmetry, neutrino mass models, L-R symmetry, etc. As part of Phase-1 of the MoEDAL-MAPP project, the CERN Research Board also approved,  in December 2021,  the installation of the MoEDAL Apparatus for Penetrating Particles (MAPP) in the UA83 tunnel some 100~m away from IP8. Installation of MAPP is already underway. The MAPP detector expands the physics reach of MoEDAL to include sensitivity to feebly interacting particles with charge, or effective charge, as low as 10$^{-3}e$ (where $e$ is the electron charge). Additionally, the MAPP detector in conjunction with MoEDAL's trapping detector gives us a unique sensitivity to extremely long-lived charged particles. MAPP also has some sensitivity to long-lived neutral particles. In Phase-2 of the project we will install the MAPP-2 upgrade to the MoEDAL-MAPP experiment for the HL-LHC. We envisage that this detector will be deployed in the UGC1 gallery near to IP8. This phase of the experiment is designed to ensure that  MoEDAL-MAPP's sensitivity to very long-lived neutral messengers of physics beyond the Standard Model is competitive and complementary to other planned projects in this arena. Overall, the MoEDAL-MAPP project  provides a largely complementary expansion of the LHC's discovery horizon. Initial  MoEDAL plans for the MEDICI facility at the 100~TeV  FCC-hh machine will also be briefly presented. 
\end{abstract}


\maketitle
\snowmass
\newpage

\section{Introduction}
\noindent
The discovery of the Higgs boson announced in 2012 by the LHC's large general purpose  detectors, ATLAS and CMS, put in place the last piece of the Standard Model (SM) puzzle.  Unfortunately,  evidence for physics beyond the SM is still not forthcoming at the LHC. A logical possibility is that conventional collider detectors are not optimized to detect  the new physics present.  MoEDAL   \cite{MoEDAL-TDR,Acharya-2014} is the first of a series of ``dedicated experiments'' planned for the LHC that is designed to provide a complementary coverage for messengers of physics beyond the SM  that  the  ATLAS, CMS and LHCb detectors might  find it more challenging,  if not impossible, to detect. This could be the case if the avatars of any new physics were highly ionizing (HIPs), feebly interacting (FIPs) or very long-lived weakly interacting neutral particles (LLPs). The Letter of Intent (LoI) for the MoEDAL-MAPP experiment --- designed to search for all three avatars of new physics: HIPs, FIPs and LLPs --- has been endorsed by the LHCC. 

\noindent
In December 2021 the CERN Research Board approved the MoEDAL Apparatus for Penetrating Particles (MAPP) \cite{MAPP-TP}, a Phase-1 upgrade to the MoEDAL detector, that will provide competitive sensitivity to FIPs \cite{Staelens-2021}. The MAPP detector began installation in UA83 adjacent to IP8 on the LHC ring in December 2021. Additionally, the MAPP detector in conjunction with MoEDAL's trapping detector gives us a unique sensitivity to extremely long-lived charged particles. MAPP also has some sensitivity to long-lived neutral particles that is currently under study. The baseline MoEDAL detector will be updated and reinstalled to take advantage of the roughly five-fold increased in luminosity expected at IP8 as well as the increased centre-of-mass energy ($E_\text{CM}$). Preparation for the MoEDAL reinstallation has already started. Full installation awaits a green-light from the LHCb collaboration which MoEDAL shares the IP8 intersection point.

\noindent
 Phase-2 of the MoEDAL-MAPP project will see the implementation of the MAPP-2 detector in the UGC1 gallery for High Luminosity LHC (HL-LHC) running  some 26~m to 55~m from IP8,  in addition to the  MoEDAL and MAPP-1 detectors.  MAPP-2 will take advantage of a  greatly enhanced decay zone volume and enhanced luminosity to push the search for LLPs. In the final part of this white paper we present some estimates of the sensitivity of a MoEDAL-MAPP type detector we call MEDICI for HIPs and FIPs,  at the Future Circular Collider (FCC).

 \section{The MoEDAL Detector} 
\begin{figure}[hbt]
\centering\includegraphics[width=0.6\linewidth]{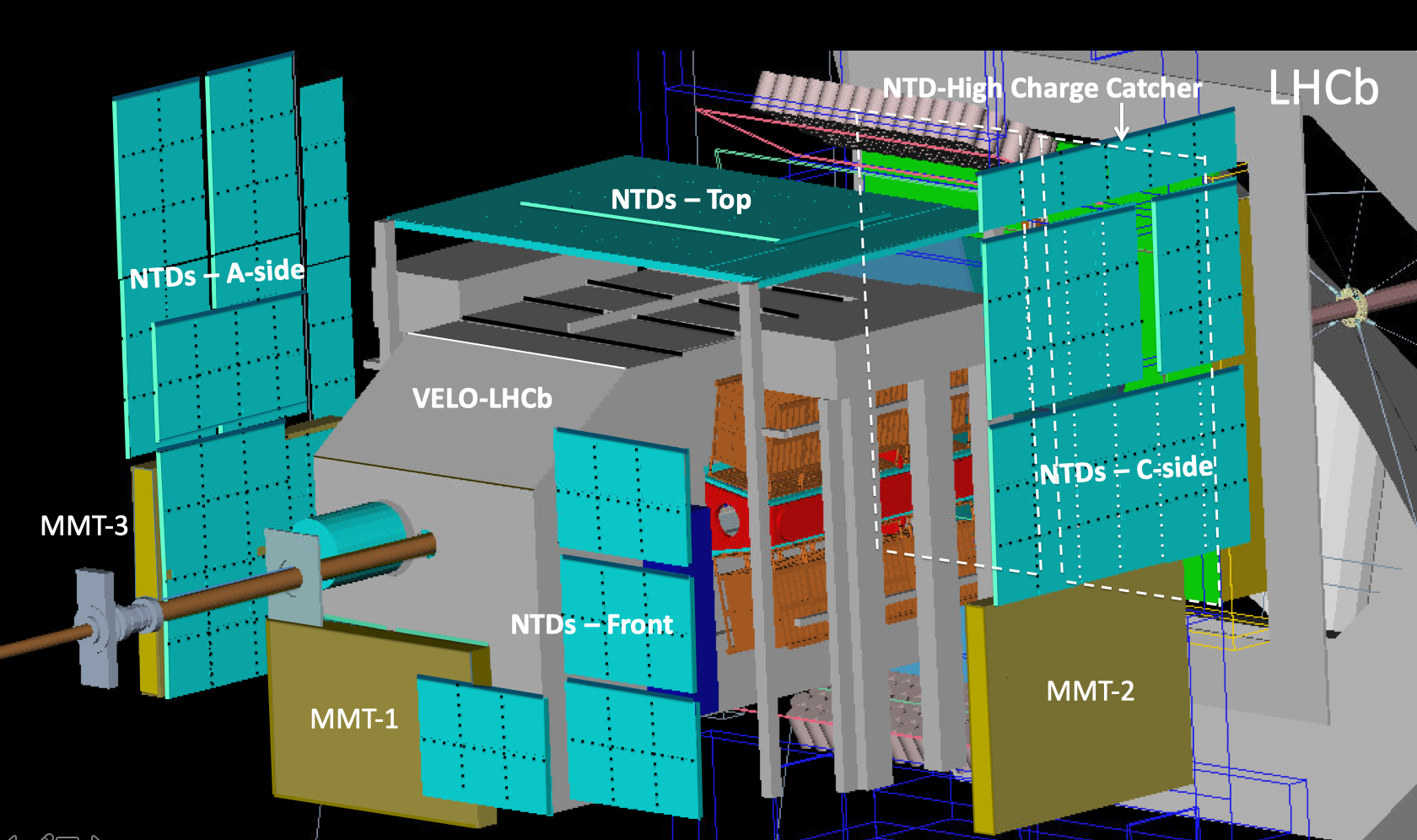}
\caption{A schematic view of the MoEDAL detector deployed in the VELO cavern at IP8.}
\label{fig:moedal-detector}
\end{figure}

 \noindent
MoEDAL  officially started  data taking in 2015 at a $E_\text{CM}$ of 13~TeV. The MoEDAL detector, positioned at Interaction Point~8 (IP8),  is totally different to other collider detectors. It currently comprises  approximately 155~m$^{2}$ of plastic of Nuclear Track Detectors (NTDs) that can track HIPs and accurately measure their charge. Additionally,  MoEDAL has roughly 800~kg of trapping volumes, forming the MMT sub-detector, that can capture HIPs for study in the laboratory. In this case the aluminium trapping volumes comprising the MMT are scanned for the presence of MMs using the SQUID Magnetometer at the ETH Zurich Laboratory for Natural Magnetism. Both MoEDAL detector systems would retain a direct and permanent record of discovery with no appreciable SM  physics backgrounds. Importantly, the NTD and MMT detectors can easily be calibrated using heavy ions and test coils, respectively.  The  MoEDAL baseline detector is designed to only detect HIP avatars of new physics and is insensitive to SM physics signals. MoEDAL's  passive detector technology  allows it to operate without gas, electrical power, readout or trigger. A schematic view of the MoEDAL detector provided by Geant4's \emph{Panaromix} program is shown in Figure~\ref{fig:moedal-detector}.

\subsection{The MoEDAL-Physics Program for LHC's  Run-1 and Run-2}\label{sc:phys}
\noindent
The founding motivation for the MoEDAL  experiment at the LHC is to pursue the quest for the magnetic monopole (MM) and the dyon \cite{Dirac1931kp,Diracs_idea,tHooft1974qc,Polyakov1974ek,Julia1975ff,Nambu1977,Witten1979ey,Lazarides1980cc,Sorkin1983ns,Gross1983hb,Cho:1997,Cho:2005,Schwinger1969ib,Hung:2020vuo,Ellis:2020bpy,Preskill1984gd,AV,Kephart2006zd,CHO-KIM-YOON,ZHANG,Rajantie2005hi} at the TeV scale. MoEDAL  is  also designed to   search for other HIPs such as massive, stable or very long-lived, slow-moving particles \cite{Fairbairn2007} with single charge,  small multiple electric charges and Highly Electrically Charged Objects  (HECOs) that arise in many scenarios of physics Beyond the Standard Model (BSM) \cite{Niles1984,Haber1985,Drees1993,Kane1994,Ellis1996,Ellis1997,Baer1996,Baer1998,Barger1998,Baer2001,Barbieri2005,Harnik2004,Nomura2005,Arkani1998,Antoniadis1998,Arkani1999,Randall1999,Antoniadis1990,Antoniadis1994,Antoniadis1994A,Appelquist2001,Ellis2000,Ellis2004,Shiu2004,Morrissey2009}.

\noindent
Magnetic monopoles that carry a non-zero magnetic charge and dyons possessing both magnetic and electric charge are among the most fascinating hypothetical
particles and probably the most important particles yet to be discovered~\cite{Patrizii:2015uea,Mavromatos:2020gwk}. In 1931 Dirac formulated a consistent description of a magnetic monopole \cite{Dirac1931kp} 
 within the arena  of quantum mechanics. This monopole is associated with an infinitely long infinitely thin solenoidal  singularity called a Dirac string. The Dirac  Quantization 
 Condition (DQC) was derived  from  the condition that the Dirac string has no physical effect:
 \begin{equation} 
g = ng_{D} = \frac{2\pi\hbar}{\mu_{0}e}n, \quad \textnormal{in SI units of Ampere-metres},
\end{equation}
where $e$ is the fundamental  electric charge of the particle, $\hbar$ is Planck's constant divided by 2$\pi$, $g_{D}$ is the magnetic charge, $\mu_{0}$ is the permeability of free space and $n$ is an integer. 

\noindent
The DQC shows that if magnetic charge exists then consequently electric charge is quantized in units of $e = 2\pi\hbar/(\mu_{0}g_{D})$. The value of $g_{D}$ is roughly 68.5$e$.  Dirac's theory does not predict the mass or the spin or charge number ($n$)  of the MM. Furthermore, the DQC  coupling strength  is  much bigger than one: $\alpha_{m}= \mu_{0}g_{D}^{2}/(4\pi\hbar c) \approx 34$. 
Thus, perturbation theory cannot be fully trusted in cross-section calculations. However, these are useful as a benchmark to allow comparison with other experiments.

\noindent
In 1974 't Hooft \cite{tHooft1974qc} and Polyakov \cite{Polyakov1974ek} revealed the existence of MM solutions of the non-Abelian Georgi-Glashow model~\cite{georgi:1972}. This model has a three-component Higgs field with only one  gauge symmetry, $SO(3)$. The predicted 't Hooft-Polyakov MM mass was  around 100~GeV/$c^{2}$, but such a low mass   was ruled out by experiment.  Eventually, using  the single non-Abelian gauge symmetry, $SU(5)$, Georgi and Glashow combined their electroweak theory with a theoretical description of the strong force in order to create a Grand Unified Theory (GUT) \cite{georgi:1974} . In this GUT theory the MM has a mass of  around 10$^{15}$ GeV/$c^{2}$, much too heavy to be directly produced at any possible  terrestrial collider. 

\noindent
The  $SU(2)\times U(1)$ group structure of the SM  does not admit  a finite-energy monopole.
However,  Cho and co-workers have modified the SM's  structure  to admit the possibility of an ``electroweak'' monopole \cite{Cho:1997,Cho:2005} with a magnetic charge of $2g_{D}$.  Based on this work, Cho, Kim and Yoon (CKY) \cite{CKY:2015,Cho:2019vzo} have more recently presented an adaptation of the SM ---  including  a  non-minimal  coupling  of  its Higgs field to  the  square  of  its  $U(1)$  gauge  coupling  strength --- that permits the possibility of a finite-energy dyon \cite{Schwinger1969ib}. 

In a model of non-sterile right-handed neutrinos, Hung~\cite{Hung:2020vuo} has shown the existence of a MM solution in a form of a topologically stable soliton with finite energy whose mass ranges at the electroweak scale from 900~GeV$/c^2$ to 3~TeV$/c^2$. In a subsequent paper, Ellis, Hung and Mavromatos predicted~\cite{Ellis:2020bpy}, from the electroweak monopole model, a value for the weak mixing angle consistent with experiment. The mass of this monopole is ``low'' enough to be accessible at the LHC.

\noindent
Even though there is no generally acknowledged empirical evidence for the MMs existence, there are strong theoretical reasons to believe that they do exist, and
they are predicted by many theories including GUTs and superstring theory.

\noindent
The laws of electrodynamics guarantee  the stability of the lightest magnetic monopole and since MMs strongly interact with the electromagnetic field they are easily detected experimentally. The scattering processes of magnetic monopoles with electrically charged fermions 
are even at low energies \cite{MMscatter1,MMscatter2} dependent on their microscopic properties. They therefore provide 
a unique window on BSM physics at high energies.  In particular they would elucidate some of the most fundamental aspects
of electrodynamics, for example its relation to other elementary particle interactions.

\begin{figure}[hbt]
\centering\includegraphics[width=0.6\linewidth]{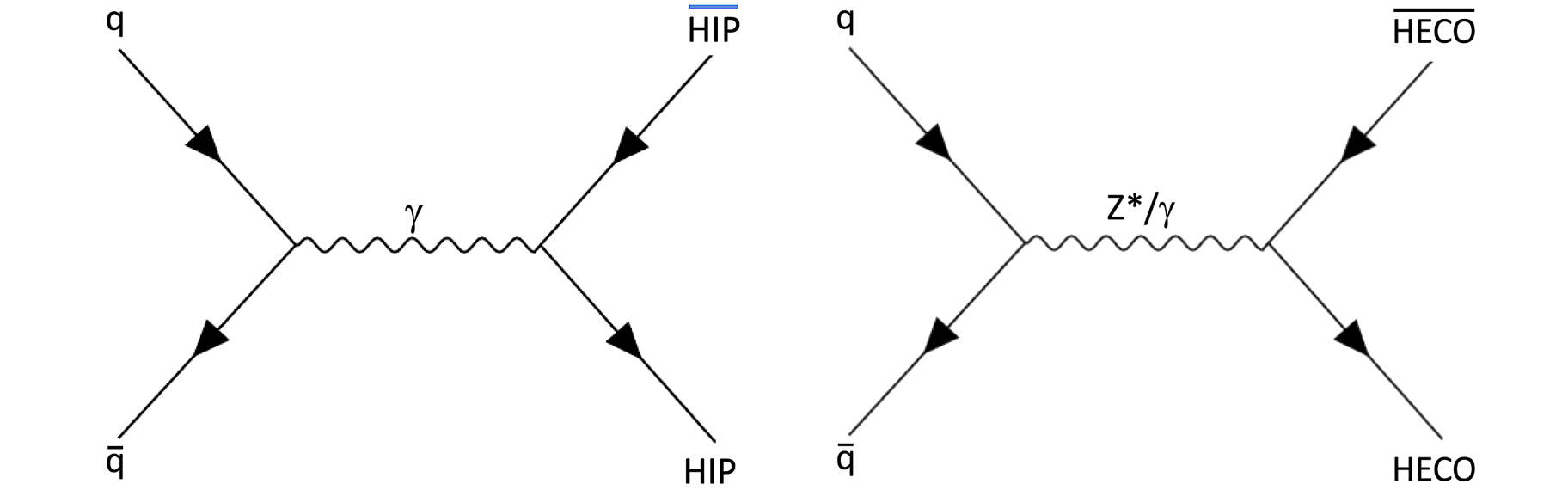}
\caption{Tree level Feynman diagram for DY production (left) of HIP anti-HIP pairs and (right) spin-1/2 HECO pairs.}
\label{fig:Feynman-HECOs}
\end{figure}

\subsection{MoEDAL LHC Run-2 Results - $pp$ Collisions}
\begin{figure}[hbt]
\centering\includegraphics[width=0.6\linewidth]{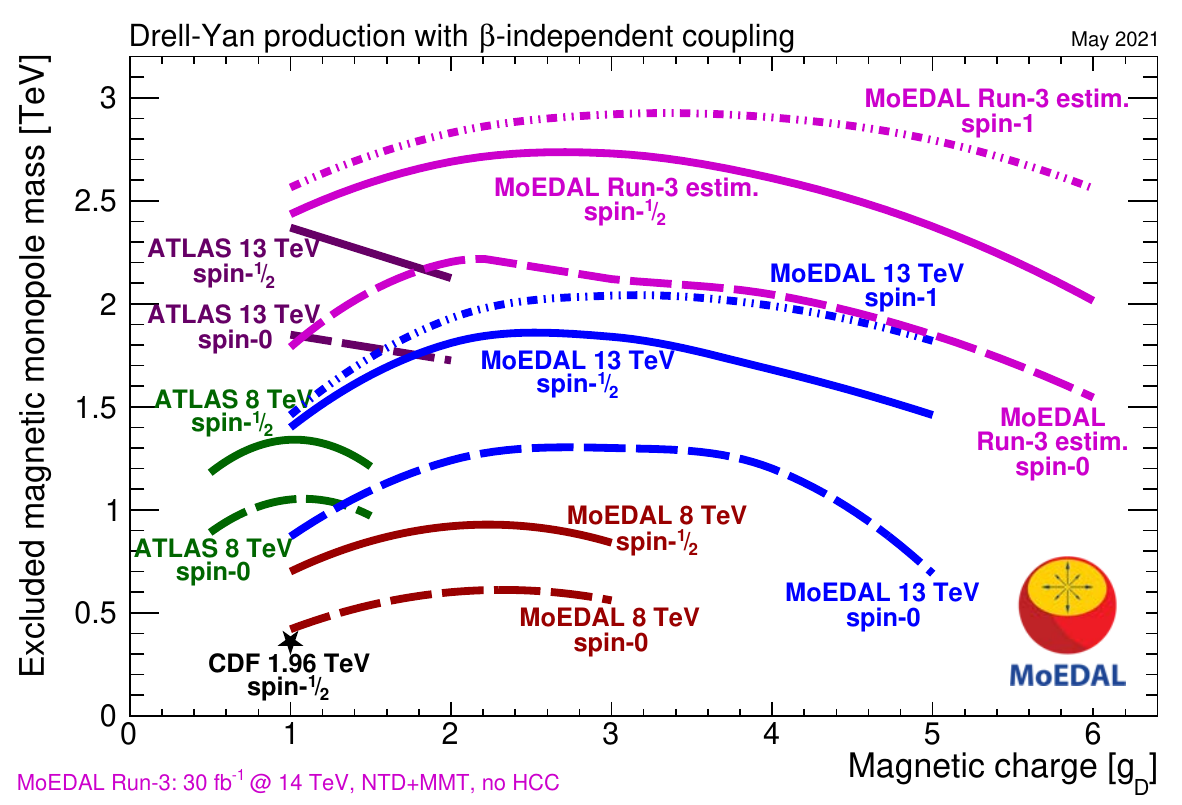}
\caption{Mass limits obtained by ATLAS and MoEDAL monopole  searches via DY pair production~\cite{VAM-HEP2022}.}
\label{fig:moedal-limits}
\end{figure}

\begin{figure}[hbt]
\centering\includegraphics[width=0.6\linewidth]{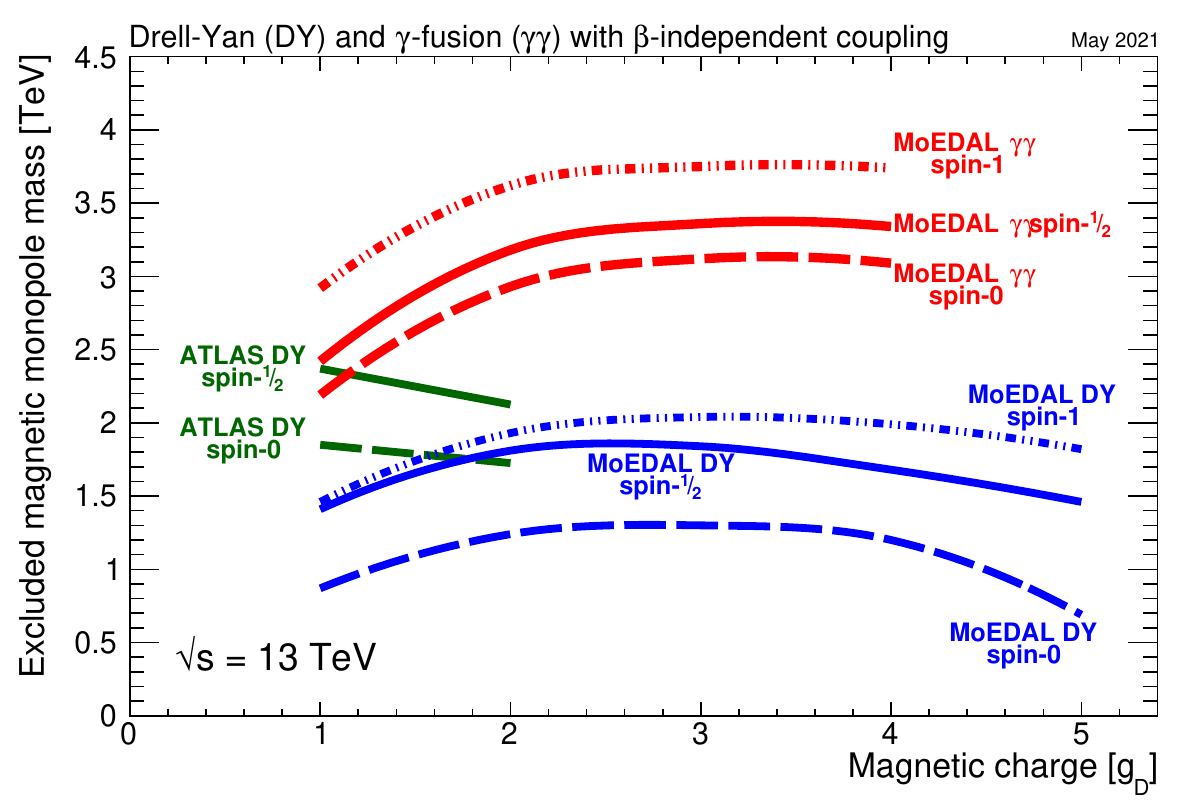}
\caption{Mass limits obtained by  MoEDAL's  monopole  search utilizing the photon-fusion production mechanism~\cite{VAM-HEP2022}.}
\label{fig:photon-fusion-limits}
\end{figure}

\noindent
 To date, we have already placed the world's best limits on singly and  multiply  charged monopole production, using detector technology that is  \emph{directly sensitive} to the magnetic charge of the monopole.  Importantly,  MoEDAL has extended the search to include spin-1 monopoles and  monopole production via photon-fusion, for the first time at the LHC.  The results of the ATLAS and MoEDAL searches for the monopole are  given in detail in Refs.\ \cite{moedal1, moedal2, moedal3, moedal4}. A summary plot showing the limits placed by LHC experiments in the search for MM production via Drell-Yan  (DY) pair production, is given in Figure~\ref{fig:moedal-limits}. The plot includes the limits MoEDAL expects to be able to place on MMs  using the expected 30~fb$^{-1}$ of Run-3 $pp$  data.  The  mass limits  obtained from MM searches obtained utilizing photon production compared to DY pair production are given in Figure~\ref{fig:photon-fusion-limits}. As can be seen the photon-fusion production process has a much higher cross-section than DY at LHC energies~\cite{Baines:2018ltl}.  All MoEDAL results are for spin-0, spin-1/2 and spin-1 monopoles. Overall, currently MoEDAL's mass limits on MMs are currently the strongest in the world for all spin-types.
 
\begin{figure}[hbt]
\centering\includegraphics[width=0.6\linewidth]{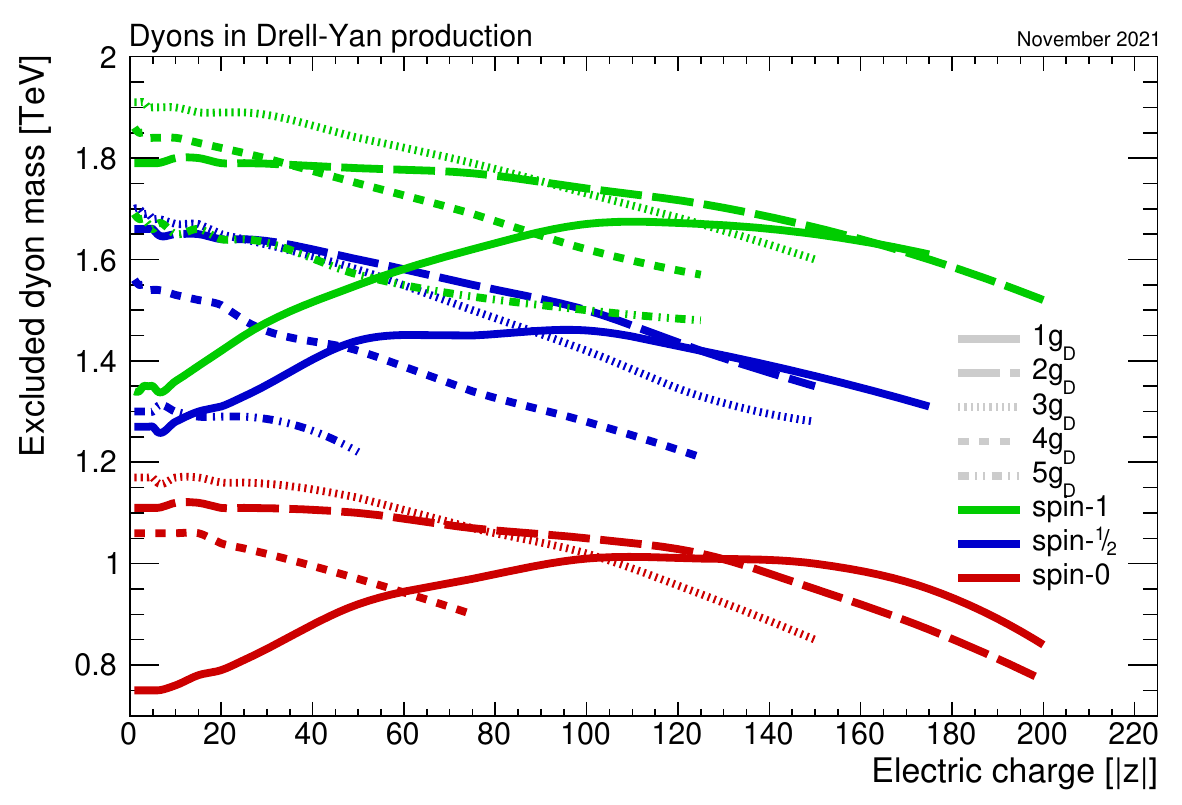}
\caption{Dyon mass limits for various electric charges, magnetic charges and dyon spins~\cite{VAM-discrete2021}. The dyons are assumed to be pair-produced via DY production.}
\label{fig:dyons-DY-production}
\end{figure}

\noindent
The MoEDAL experiment has conducted the first direct search for the DY production of spin-0, spin-1/2 and spin-1  dyons, where the dyon is  an hypothetical particle with electric and magnetic charge first predicted by Schwinger in 1969  \cite{Schwinger1969ib} \cite{moedal5}.  We have excluded  dyon production cross sections as low as 30~fb and placed mass limits of  750-1910~GeV$/c^2$ for dyons  with  up to $5g_{D}$ and electric charges in the range $1e$ to $200e$, where $e$ is the electron charge.  Dyon mass limits obtained for various electric charges, magnetic charges and dyon spins are shown in Figure~\ref{fig:dyons-DY-production}.

Using 2.2~fb$^{1}$ of Run-1 data with a ($E_\text{CM}$) of 8~TeV MoEDAL has also  placed the world's best charge limits on Highly Charged Electrical Objects (HECOs) \cite{moedal6} where HECOs with charges as high in the range $10e$ to $180e$ have been ruled out for HECO masses between 30~GeV/$c^{2}$ and 1 TeV/$c^{2}$.  the search for HIPs, including monopoles and HECOs, is now underway using the full MoEDAL detector and the full Run-2 luminosity. Another analysis that is currently underway is the search using a SQUID magnetometer for MMs trapped  in the  CMS central Be beam pipe, section that was donated to the MoEDAL  Collaboration in 2019 \cite{CMS-BP}.

\subsection{MoEDAL LHC Run-2 Results -- Pb-Pb  Collisions}
\begin{figure}[hbt]
\centering\includegraphics[width=0.5\linewidth]{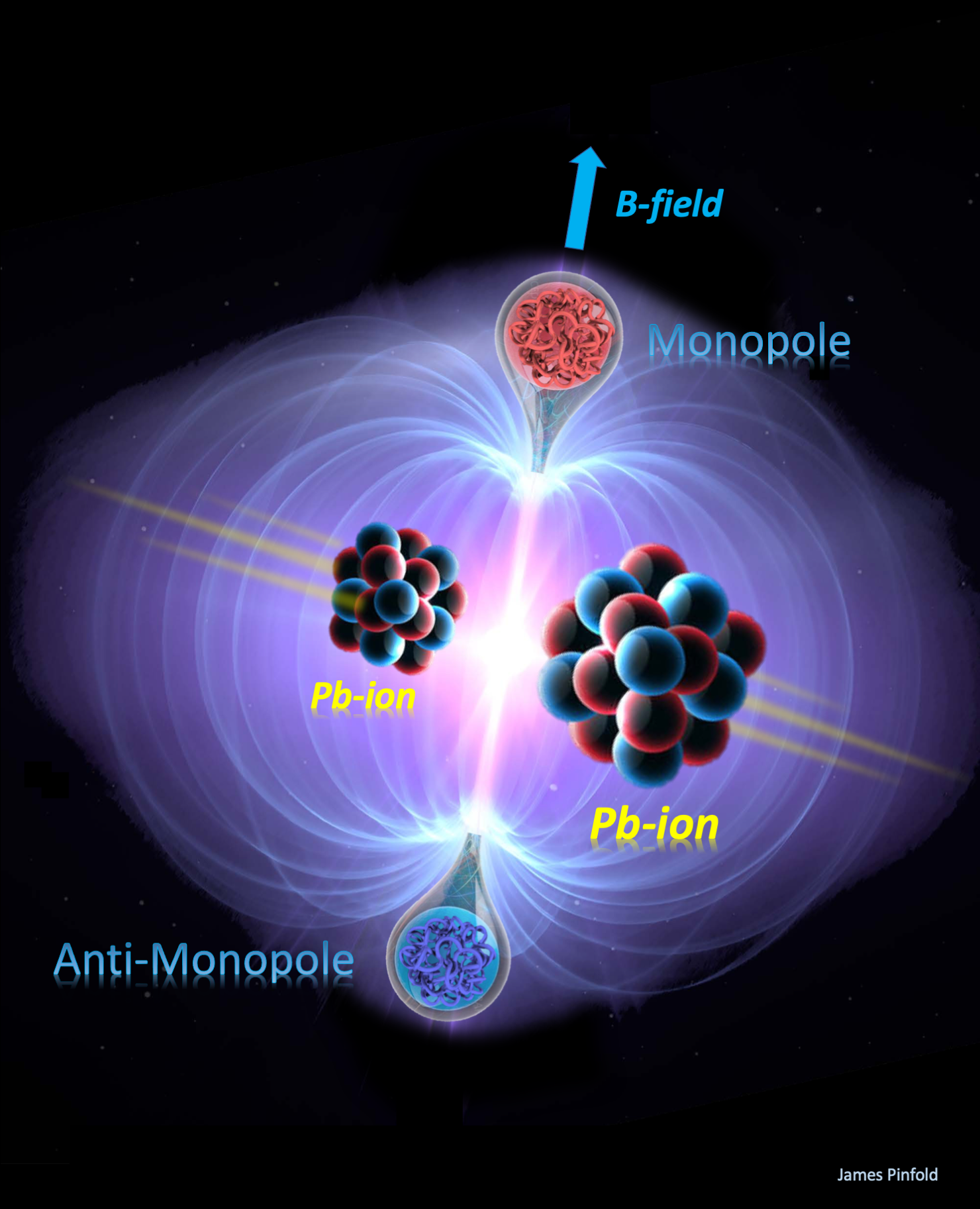}
\caption{Visualization of pair production of a MM Anti-MM pair production in the intense magnetic fields created by the ultra-peripheral collision of two lead nuclei at LHC scale energy.}
\label{fig:Schwinger-mechanism}
\end{figure}

\noindent
To date, MM searches at particle accelerators have studied monopole  production via the Drell-Yan mechanism or via photon-fusion. However, the magnitude of the MM's coupling makes physically valid perturbative cross-section calculation impossible.  Another issue is that the production of composite MMs from elementary particle collisions should be exponentially suppressed due to the physical size of the MMs making all previous searches only sensitive to point-like MMs. The search for MMs via the Schwinger mechanism provides a way to avoid these drawbacks. In 1951, Schwinger showed that pairs of electrically charged particles could be created in a very strong electric field \cite{Schwinger-1951}. If MMs exist, electromagnetic duality implies that MM pairs would also be produced via the same mechanism in an intense magnetic field. Schwinger production can be interpreted as quantum tunnelling through the Coulomb potential barrier. An artist's impression of this process is given in Figure~\ref{fig:Schwinger-mechanism}.

The magnetic field strength requirement  for MM pair production via the Schwinger mechanism is satisfied by the transitory magnetic fields of unprecedented strength produced in ultraperipheral heavy-ion collisions at the LHC~\cite{B-fields}.  During the LHC's Pb-Pb run in 2018, the estimated peak magnetic field strength was $B\approx10^{16}$~T \cite{B-fields}. This is roughly 10,000 times greater than the strongest known magnetic fields in the Universe on the surface of a special kind of neutron star called a magnetar \cite{magnetar}.  

In 2021, for the first time ever,  MoEDAL searched for  MM pair production via the Schwinger mechanism~\cite{MoEDAL-2022}. 
MoEDAL utilizes two approaches in the calculation of the overall MM production cross section \cite{Gould-2021, MoEDAL-2022}. The first is the free-particle approximation (FPA) where MM self-interactions are ignored but the spacetime dependence of the heavy-ion's EM field is handled exactly. In the second approach, called the locally constant field approximation, the MM self-interactions are treated exactly but spacetime dependence of the EM field is neglected. Both approaches are expected to yield conservative lower limits. MoEDAL  uses the smaller cross section in determining its final MM mass bounds.

In order for MoEDAL to search for Schwinger mechanism produced MM pairs in Pb-Pb collisions at the LHC, we need to calculate the momentum distributions of the MMs produced, which at the TeV energy scale depend primarily upon the time dependence of the electromagnetic (EM) fields involved. To calculate these distributions, MoEDAL used the FPA approach.      

\begin{figure}[hbt]
\centering\includegraphics[width=0.8\linewidth]{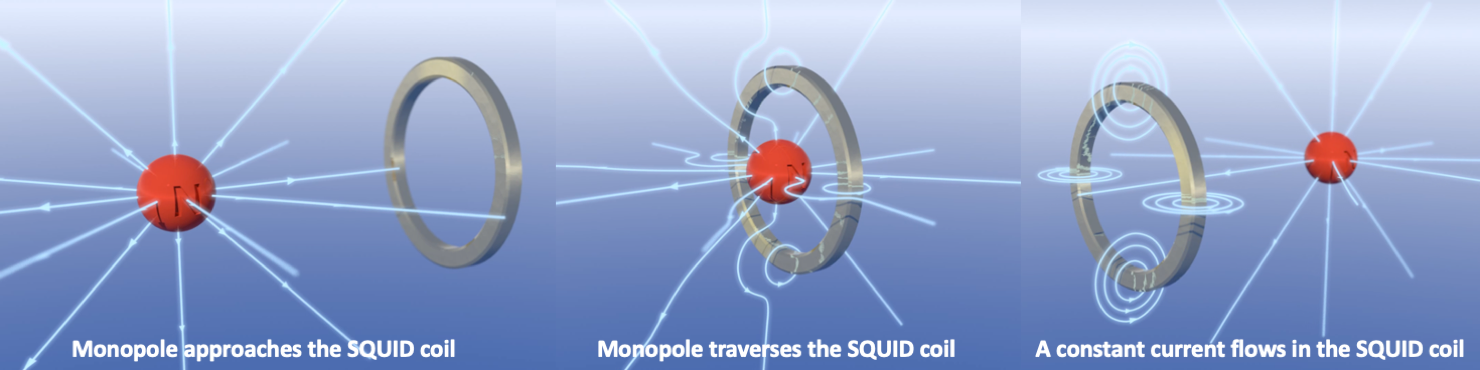}
\caption{The passage of a MM through the SQUID coil creates a constant current with magnitude dependent on magnetic charge.}
\label{fig:squid-response}
\end{figure}

A fraction of the highly-ionizing MMs produced are trapped in the MMT. After exposure during Run-2 MoEDAL's  MMT trapping volumes are scanned for the presence of trapped MMs using the SQUID Magnetometer at the ETH Zurich. The signature signal is a continuing current in the SQUID coil that is created by the passage of a monopole through the coil, as shown in Figure~\ref{fig:squid-response}. No SM particle could give this signal. 

\noindent
The search for MM pair production via the Schwinger Mechanism was carried out using data taken during  November 2018 by the MoEDAL experiment using approximately $1.8\times$10$^{9}$ of Pb-Pb collisions with a centre-of-mass collision energy of 5.02~TeV \cite{MoEDAL-2022}. Unfortunately, the presence of a MM signal was not detected in the data taken. Thus, we were able to exclude, at the 95\% confidence level, the existence of monopoles lighter than 75~GeV/$c^{2}$ for magnetic charges ranging from $1g_{D}$ to $3g_{D}$. 

\noindent
These limits are less impressive than those already placed on MM production via the DY and photon fusion production mechanisms at the LHC to date, since the  Schwinger MM production cross-section depends exponentially on the MM mass. However, this search is truly ground breaking for two reasons.  First, previous cross section calculations and limits placed on MM production are unphysical as the MM's coupling constant is so large as to make perturbative calculations unreliable. But, such calculations for MM production via the Schwinger mechanism are theoretically sound. Secondly, this may well be the first direct search that is sensitive to finite-sized MMs since they are produced via the Schwinger mechanism with cross-sections that are not exponentially suppressed as expected in DY or photon-fusion production.

\section{The MoEDAL-MAPP Detector for Run-3}
\begin{figure}[hbt]
\centering\includegraphics[width=0.8\linewidth]{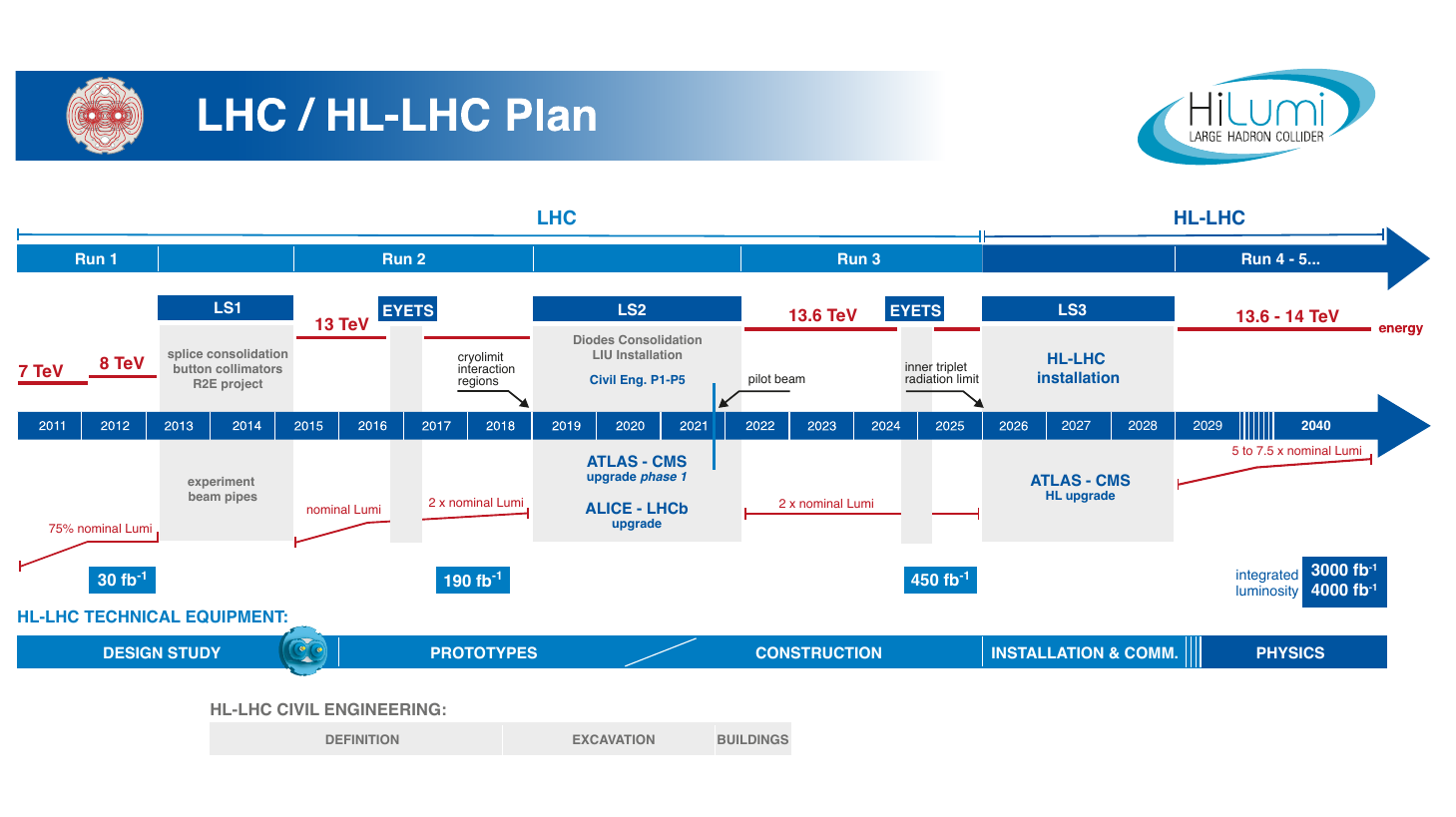}
\caption{A overview of the full LHC program up to and including HL-LHC.}
\label{fig:LHC-program}
\end{figure}

\noindent
The MoEDAL collaboration is  here requesting to take and additional 25~fb$^{-1}$ to 30~fb$^{-1}$ of data during Run-3 of the LHC program which is  depicted in Figure~\ref{fig:LHC-program}. This is possible due to LHCb's planned  increase in the collision rate at IP8 giving a factor of $\sim$5 increase in instantaneous luminosity over  Run-2.  Also, MoEDAL will be able to access  the higher $E_\text{CM}$ of 14~TeV. The higher energy and enhanced statistics  will allow an expanded search for  HIPs including, e.g., the search for electrically charged  massive (pseudo-)stable particles from, supersymmetric and extra-dimension scenarios.  To advance our existing program at LHC's Run-3  we would redeploy the existing MoEDAL detector in the VELO/MoEDAL cavern in its current form, with some minor changes.

\noindent
A major part of MoEDAL's Run-3 upgrade program will be the addition to the baseline MoEDAL detector of the new MAPP sub-detector. MAPP's purpose is to expand the physics reach of MoEDAL to include the search for: milli-charged particles (mCPs) with charges as low as one thousandth the electron charge; and weakly interacting very long-lived neutral particle messengers of new physics. Thus, the MoEDAL and MAPP detectors operating together will be able to detect: HIP, mCP and LLP avatars of new physics. The MAPP-mQP detector can also be used to monitor exposed trapping volumes from MoEDAL's MMT detector placed underneath the MAPP-mQP detector for the decays of captured Long Lived Charged Particles (LLCPs).  We can rapidly transfer MoEDAL's exposed trapping detectors underground to the MAPP-mQP detector in UA83, in order to maximize the range of lifetimes to which this arrangement is sensitive.

\subsection{The Baseline MoEDAL Detector at Run-3}
\noindent
 The MoEDAL detector is currently being updated to continue the search for MMs --- at higher energy and at much greater luminosity ($\sim$30~fb$^{-1}$) ---  at LHC's Run-3 using $pp$ and heavy-ion collisions.   The MoEDAL detector will be redeployed at IP8 for data taking at Run-3 with some upgrades to the NTD detector geometry to improve the efficiency of the NTD detector system and the full use of the low threshold CR39 detectors, that were calibrated during Run-2. Another  improvement to the MoEDAL detector  for Run-3 and HL-LHC data taking will be  the use of PET NTD detectors in the NTD stacks alongside Makrofol and CR39. The PET threshold is sufficient to detect magnetic monopoles with magnetic charge  2$g_{D}$ and above, with much less sensitivity to LHC beam induced backgrounds~\cite{Bhattacharyya:2016swk}. 

\noindent
Additionally, we expect to maintain, or even improve,  the signal to noise ratio of the NTD detectors despite  increased beam backgrounds due to enhanced luminosity.  This is achieved by the use of Machine Learning techniques that are currently under study by the MoEDAL Machine Learning group.  Another, contribution in this arena is the use of upgraded automatic optical scanning tables, based at the University of Helsinki,  with multiple illumination and camera options. In this way we can also expect to ameliorate the effect of beam background induced noise in the NTD films as well as increase the speed with which we scan the NTD films.

\noindent
In addition, we will update the TPX array by replacing TimePix2 pixel devices with TimePix3 chips that can be used with the LHC clock to better understand the radiation environment in the VELO cavern. In addition,
 we anticipate that the ability of the TimePix3 chip to track multiple closely spaced particles~\cite{moedal-timepix} may be useful in the detection of new physics scenarios involving large numbers of softer tracks.

\begin{figure}[hbt]
\centering\includegraphics[width=1.0\linewidth]{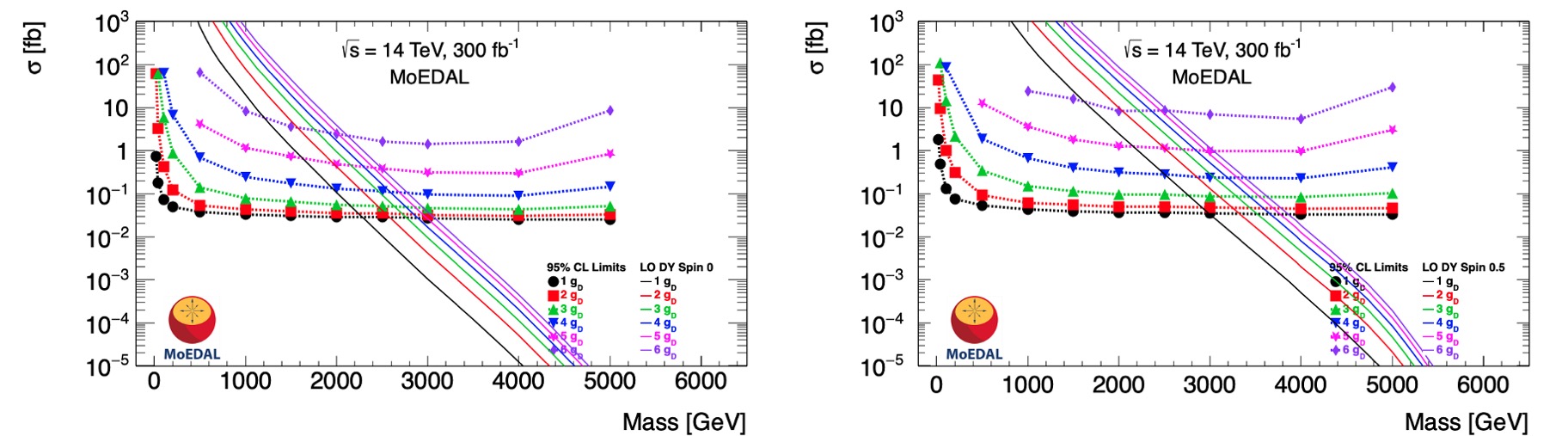}
\caption{The estimated  95\% CL limit curves based on the non-observation of MM production via the DY mechanism at the HL-LHC.}
\label{fig:HL-LHC-MM}
\end{figure}

\subsection{Searching for  MMs in $pp$ Collisions  at Run-3 and Beyond Using the MoEDAL Detector}
\noindent
The search for HIPs will continue at Run-3 and beyond utilizing the expected factor of five enhancement in luminosity at IP8 and the increase of $E_\text{CM}$ to 14~TeV. As is shown in Figure~\ref{fig:moedal-limits}, using the Run-3 dataset with a nominal luminosity of 30~fb$^{-1}$, we expect to be able to push the search for MMs up to masses in the range 1.5~TeV/$c^{2}$  to  2.9~TeV/$c^{2}$, depending on the MM charge and  spin.  In the case of photon-fusion production of MMs we envisage a sensitivity to MM masses as high as 5~TeV/$c^{2}$ at Run-3. Additionally, we completed an estimate of what mass limits we can approach after 300~fb$^{-1}$ of data taking at the HL-LHC. The corresponding 95\% limit curves,  assuming no MMs are observed, for DY produced spin-1 and spin-1/2 monopoles are shown in Figure~\ref{fig:HL-LHC-MM}. If a magnetic monopole is not observed at the HL-LHC we can place lower mass limits of MMs in the range 2.1~TeV/$c^{2}$ to 3.4~TeV/$c^{2}$ with the corresponding mass limits from photon fusion a few TeV higher.

\subsubsection{Searching for  MMs in Pb-Pb Collisions  at Run-3 and Beyond}
\begin{figure}[hbt]
\centering\includegraphics[width=0.6\linewidth]{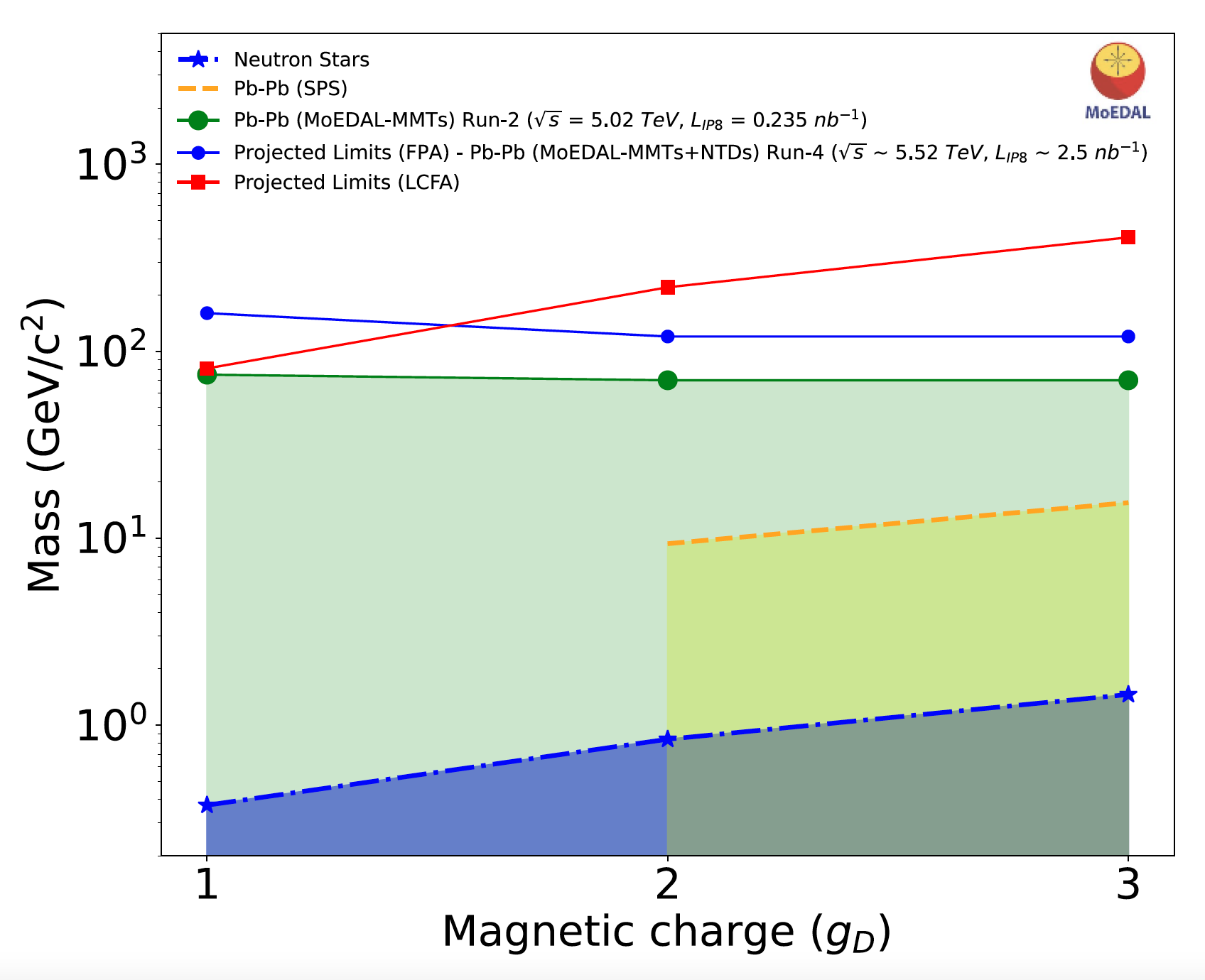}
\caption{The estimated  95\% CL limit curves based on the non-observation of MM production via the Schwinger mechanism using Pb-Pb collisions using data taken in LHC Runs 2, 3 and 4 at the LHC~\cite{dEnterria:2022sut}.}
\label{fig:PbPbRun4}
\end{figure}

\noindent
 The projected limits for MMs produced by Schwinger mechanism in heavy-ion collisions for future LHC Runs-3 and 4, including Run-2, are given in Figure~\ref{fig:PbPbRun4}~\cite{dEnterria:2022sut}. It is assumed here that only MoEDAL's MMT detectors are utilized for the search. These projections are based on an expected increase in luminosity at LHCb IP8, from 0.235~nb$^{-1}$ during Run-2 to 2.5~nb$^{-1}$ by the end of Run 4; and collision energies from 5.02~TeV/n to 5.52~TeV/n in Run 3 and 4 \cite{PbPbRun4}. These limits will  be improved significantly with the use of MoEDAL's NTD and MMT  detectors in the search, as is planned for Run-3 and beyond.
 
 \subsubsection{ Searching for HECOS and Massive Long-Lived Singly and Multiply Electrically  Charged Particles at Run-3}
 \noindent
 As we have seen above MoEDAL's complete NTD system consisting of Makrofol, PET and CR39 is now fully calibrated and able to track electrically charged particles with $Z\beta$ as low as $\sim$7. Thus the search for HECOs can continue at the higher $E_{\rm CM}$  and luminosity available at Run-3.  Importantly, with the reduced NTD threshold and higher detector efficiency available at Run-3 we can now track, sufficiently slow moving, singly or multiply charges massive particles.  For example, supersymmetric scenarios offer a number charged states, for example, sleptons, R-hadrons and charginos,  that are long-lived enough to reach the MoEDAL NTD detectors,  of 1~m away from IP8. Additionally, there are a number of models of  new physics that predict multiply charged particles that are detectable by MoEDAL. 
 
\noindent
\emph{Singly Charged Massive Particles --  a Benchmark SUSY Scenario,}  We focus here on the prospects of directly detecting long-lived sleptons in a well predicated supersymmetric scenario model which involves an intermediate neutral long-lived particle in the decay chain. This scenario is compatible with astrophysical constraints and  is not yet excluded by the current data from ATLAS or CMS. Using Monte Carlo simulation, we compared the sensitivities of ATLAS~\cite{sleptons} and CMS~\cite{sleptons-CMS} versus MoEDAL in arenas where MoEDAL could provide a  discovery reach complementary to that of ATLAS and CMS, due to fact that MoEDAL does not require a trigger, has  looser selection criteria and, last but not least,  has  no SM backgrounds. In such scenarios, when charged staus are the main long-lived candidates, the relevant mass range for MoEDAL is compatible with a possible  supersymmetric explanation of  the anomalous events observed by the ANITA detector \cite{anita}.
 
\noindent
In  ATLAS and CMS  searches for Heavy Stable Charged Particles (HSCP), multiple hits in the (innermost) pixel detector are required to ensure good track reconstruction. Thus the presence of an undetectable \emph{neutral} long-lived particle in the decay cascade would limit the acceptance for such a scenario. 
This leads us to contemplate gluino pair production ($pp \rightarrow \tilde{g}\tilde{g}$) followed by prompt gluino decay into two quark-jets and a long-lived neutralino ($\tilde{g} \rightarrow q\overline{q}\tilde{\chi}^{0}_{1}$). We further consider the case where the long-lived neutralino may decay into a metastable stau plus  an off-mass-shell tau lepton after it has travelled around 1~m, where the lifetime of the neutralino depends on the mass mass splitting: $\Delta m = m_{\tilde{\chi}^{0}_{1}} -  m_{\tilde{\tau}_{1}} \lesssim m_{\tau}$.   The lifetime of the $\tilde{\chi}^{0}_{1}$ depends on this mass splitting. For three body decays \cite{3-bodiesa,3-bodiesb} this dependence goes as $(\Delta m)^{6}$ allowing the lifetime to be varied from around a nanosecond ($\Delta m \sim$1.7~GeV$/c^2$) to approximately a microsecond ($\Delta m \sim$0.5~GeV$/c^2$), implying decay lengths ranging from 0.1~m to 100~m. The metastable staus can decay into $\tau$'s and other SM particles via small   $R$-parity violating couplings, when accompanied by a $\tilde{\tau}$ lightest supersymmetric particle.
  
 \begin{figure}[hbt]
\centering\includegraphics[width=0.6\linewidth]{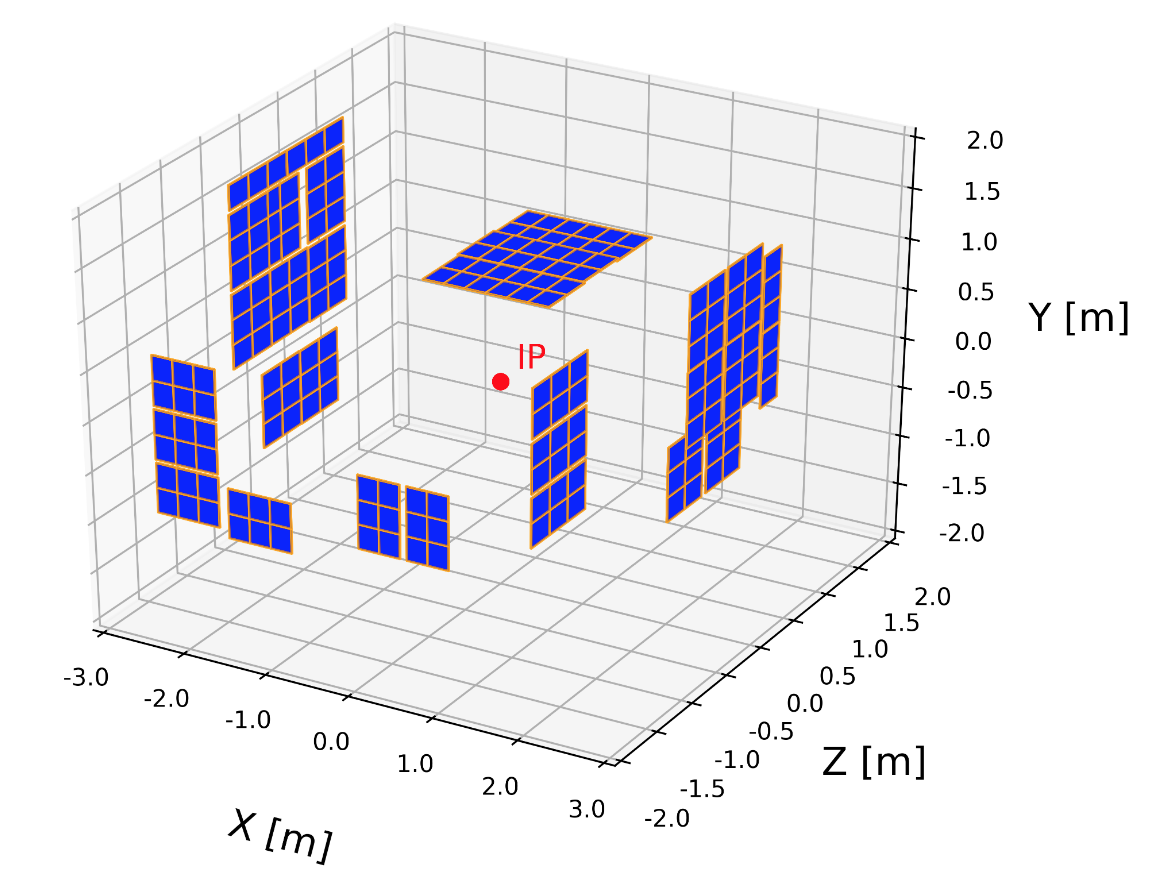}
\caption{We assume that  the Run-3 NTD deployment of MoEDAL  is the same as that  at Run-2 NTD deployment. The NTD modules are depicted as thin blue plates with orange edges. The red point at the centre represents the interaction point. The z-axis is along the beams and the y-axis indicates the vertical direction~\cite{sleptons}.}
\label{fig:NTDS-Run-2}
\end{figure}
 
\begin{figure}[hbt]
\centering\includegraphics[width=0.6\linewidth]{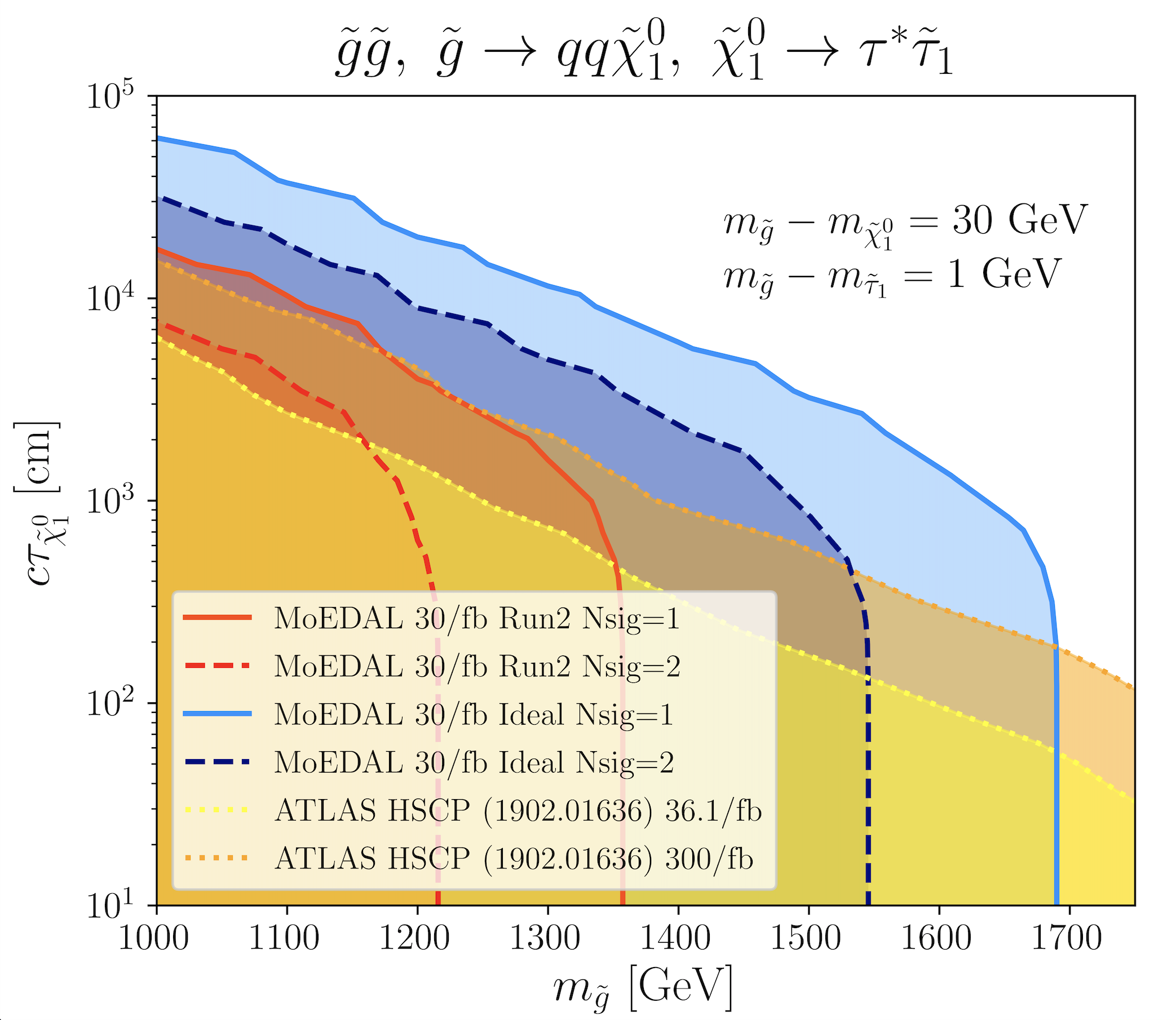}
\caption{The sensitivity of MoEDAL, where 1 signal event  N$_{sig}$ = 1,  (solid) and  two signal events N$_{sig}$ = 2 (dashed), are detected.   In the 
$\tilde{g}  \times  c\tau_{\tilde{\chi}^{0}_{1}}$ plane. The mass splitting between the gluino and the neutralino is fixed at 30~GeV/$c^{2}$ and the mass splitting between the neutralino and the stau is fixed at 1~GeV/$c^{2}$. The two MoEDAL geometries considered are that utilized at Run-2 (red)  and the geometry that we envisage will be deployed for Run-3 (blue). The sensitivity of the ATLAS HSCP analysis (52) is indicated by the yellow dashed contour. The dotted orange contour represents the projection of the ATLAS analysis from a luminosity of 36.1~fb$^{-1}$   to the expected Run-3 luminosity of 
300~fb$^{-1}$. The luminosity assumed for both MoEDAL geometries is that expected to be taken at Run-3~\cite{sleptons}.  }
\label{fig:slepton-LLPs}
\end{figure}

 \noindent
The HSCP search by CMS \cite{Khachatryan-2016}  using only 2.5~fb$^{-1}$ of $pp$ collision data at 13~TeV use very similar  analysis design and selection cuts to those of  an ATLAS search \cite{Aboud-LLP-2019} that utilized the higher luminosity of  36.1~fb$^{-1}$ available at  LHC's Run-2.  The simulation of a MoEDAL search  is based on the detection of the final state stau in MoEDAL's NTD system. The stau would only be detected if its ionizing power at the NTD is greater than detectable ionization damage  threshold of the NTD. For Run-3, the Run-2 NTD geometry  we will modified  in order  that all particles originating at the interaction point (IP) would be incident normally on the NTD stacks. Hence, we expect to detect a stau travelling slower than $\beta = v/c \approx 0.15$. We use  the Run-2 deployment of MoEDAL's NTDs shown in Figure~\ref{fig:NTDS-Run-2} and assume the mass splitting between the gluino and the neutralino is 30~GeV/$c^{2}$ and the mass splitting between the neutralino and the stau is 1~GeV/$c^{2}$. The  corresponding  sensitivity of MoEDAL's  search compared with that of ATLAS is shown in Figure~\ref{fig:slepton-LLPs}. As can be seen  MoEDAL's Run-3 geometry (called ``Ideal'') allows us to extend the search for the process described to significantly longer neutralino lifetimes.

\noindent
MoEDAL is primarily sensitive to massive  slow-moving SUSY particles ($\beta \lesssim 0.2$) in a manner that is complementary to that of ATLAS and CMS which are optimized for faster moving particles. However, for simple scenarios, the limited luminosity available to MoEDAL at IP8, which is roughly a factor of ten less than that available to ATLAS and CMS, is a limiting factor. Nevertheless, for more complex event topologies such as those with a neutral LLP in the decay chain, the example analysis discussed here indicates that MoEDAL can probe part of the parameter space that is currently unconstrained by ATLAS and CMS.

\noindent
\emph{Searching for Massive Doubly Charged Particles. } 
\noindent
Doubly-charged scalars are hypothesized to exist  in an assortment of BSM physics arenas. For example,  doubly charged scalars H$^{\pm\pm}$\cite{schlechter-1980,magg-1980,cheng-1980,lazarides-1981,mohaptra-1981,lindner-2018} can arise in Type-II seesaw models. There are  several other new physics scenarios that can give rise to doubly-charged particles, namely the:  little Higgs model \cite{Arkani-Hamed-2002}; Left-Right (L-R) model \cite{Pati-1974,Mohaptra-1975,Senjanovic-1975}; Georgi-Machacek (GM) model \cite{Georgi-1985,Chanowitz-1985,Gunion-1990,Gunion-1991,Ismail-2003}; and, the 3-3-1 model \cite{Cieza-2006,Alves-2011}.  Also, doubly charged higgsinos, can arise in supersymmetric  versions of these  models. For example,  the minimal L-R  supersymmetric model  is an example of such a model \cite{Kuchimanchi-1993,Babu-2008,Francis-1991,Huitu-1994,Frank-2014}. In addition,  multiply-charged scalars and fermions, are predicted to exist in the simplified models discussed in Refs.~\cite{Delgado-2011,Alloul-2013}.

\begin{figure}[hbt]
\centering\includegraphics[width=0.6\linewidth]{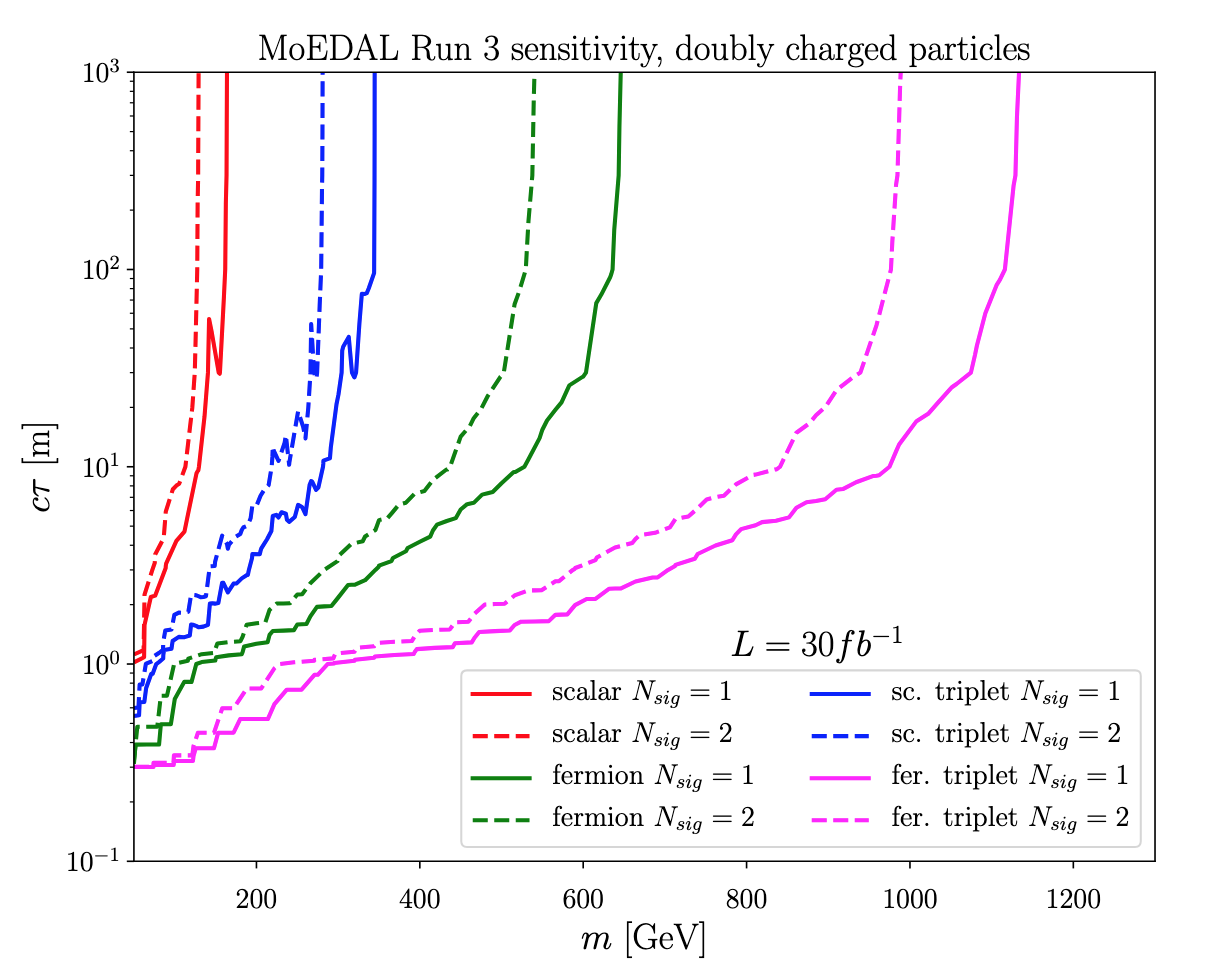}
\caption{ The envisaged MoEDAL  sensitivities for a number of  long-lived doubly charged particle species, assuming an integrated luminosity of 30 fb$^{-1}$~\cite{acharya-LLP-2021}.}
\label{fig:doubly-charged}
\end{figure}

\begin{figure}[hbt]
\centering\includegraphics[width=0.7\linewidth]{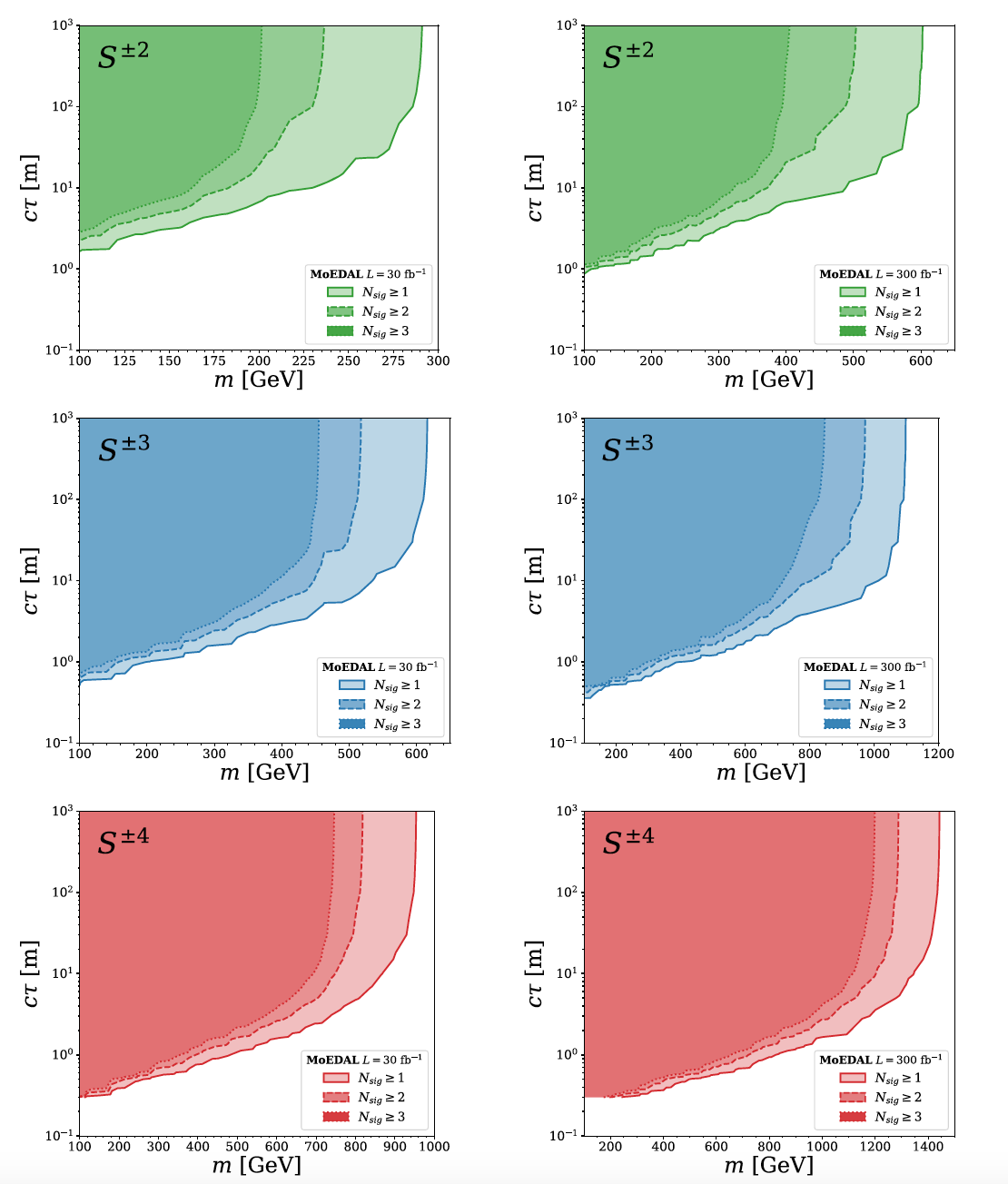}
\caption{The envisaged MoEDAL sensitivities for a number of long-lived doubly/triply/quadruply-charged particle species predicted in neutrino mass Model-1, assuming an integrated luminosity of 30~fb$^{-1}$ (Run-3)  and 300~fb$^{-1}$ (HL-LHC)~\cite{Hirsch-2021}.}
\label{fig:multiply-charged}
\end{figure}

\noindent
The  expected Run-3  MoEDAL sensitivities for the four types of colour-singlet doubly-charged particles considered \cite{acharya-LLP-2021}: a scalar singlet (red), a scalar triplet (blue), a fermion singlet (green) and a fermion triplet (magenta), are given in Figure~\ref{fig:multiply-charged}. Out of the four, the fermion triplet has the highest mass reach  due to the lower production speeds and  larger cross-section. MoEDAL is sensitive to this state up to a mass of 990(1130)~GeV$/c^2$ with a 2(1) signal particles seen if  $c\tau  \gtrsim100$~m, as shown by the solid (dashed) line. 

\noindent
 After that, MoEDAL is  most sensitive to the fermion-singlet, for which we estimate the MoEDAL mass reach, again with $c\tau \gtrsim$ 100~m,  to be approximately  650 (540) GeV$/c^2$ for  1 (2) signal particles observed. The mass reaches for the singlet and triplet  scalar particles are much lower than the corresponding fermionic particles. This is  due to their higher typical velocities and  smaller cross-sections and higher typical velocities. 
 
 \noindent
 \emph{Searching for Doubly, Triply and Quadruply  Charged Particles from Neutrino Mass Models.}
We also considered a class of radiative neutrino mass models that  predict LLPs whose electric charge is four or three times that of the electron's \cite{Hirsch-2021}. If light enough, such particles can  be produced and detected  at the LHC. 

\noindent
In particular,  MoEDAL's performance with regard to the detection of multi-charged LLPs for two particular neutrino mass  models were investigated. In this type of model the SM is enlarged to include two scalar fields $S_{1}$ (singlet) and $S_{3}$ (triplet) representations of $SU(2)_{L}$. Additionally, in $SU(2)_{L}$ there are three pairs of vector-like fermions: ($F_{i},\overline{F}_{i}$). In the first model we considered (Model-1)  all BSM fields are colour singlets. However, the second model (Model-2) the BSM fields exist in colour(anti-)triplet representations.

\noindent
 In Model-1, the BSM sector is not coloured and has three long-lived particles with electric charges $Z = $2, 3 and 4. In Model-2, the BSM sector is coloured and the long-lived particles have the fractional charges $ Z =  $ 4/3, 7/3 and 10/3. The lifetimes of the multi-charged particles in both models are controlled by three couplings $\lambda_{5}, h_{F}, h_{\overline{F}}$ as well as the BSM parity breaking coupling $f_{ee}$ (Model-1) and $h_{ed}$ (Model-2). 

\noindent
Figure~\ref{fig:multiply-charged} shows the expected sensitivities for Model-1. In the case of the scenario where only one signal event was seen in MoEDAL in Run-3 (30fb$^{-1}$ , then MoEDAL can explore particles masses up to 290, 610 and 960 GeV$/c^2$ for doubly, triply and quadruply charged particles. For HL-LHC (300~fb$^{-1}$)  these limits improve to 600, 1100 and 1430~GeV$/c^2$. The current constraints from the HSCP searches are 650, 780 and 920 GeV$/c^2$. For Model-2, MoEDAL  at Run-3 can explore masses up to 1050, 1250 and 1400 GeV$/c^2$ for charge 4/3, 7/3, 10/3, respectively. At the HL-LHC these limits are improved to 1400, 1600 and 1800~GeV$/c^2$. The corresponded estimated current bounds for Model-2 are 1450, 1480 and  1510~GeV$/c^2$ leading us to conclude that only at HL-LHC can we expect to explore previously unconstrained parameter space.

\noindent
Model-2  of this type of radiative neutrino mass models  can be obtained easily from Model-1 by simply  attributing colour (anti-)triplet charges to Model-1's   BSM fields along with some required   changes in hypercharge assignments~\cite{Hirsch-2021}. The produced multi-charged long-lived particles  hadronize into colour singlet states before decaying. After hadronization, the LLP  charge is shifted by the constituent quarks. An approximate hadronization model  was employed in Ref.~\cite{Hirsch-2021}  in order to estimate the effect of the charge shift arising from hadronization. 

\begin{figure}[hbt]
\centering\includegraphics[width=0.7\linewidth]{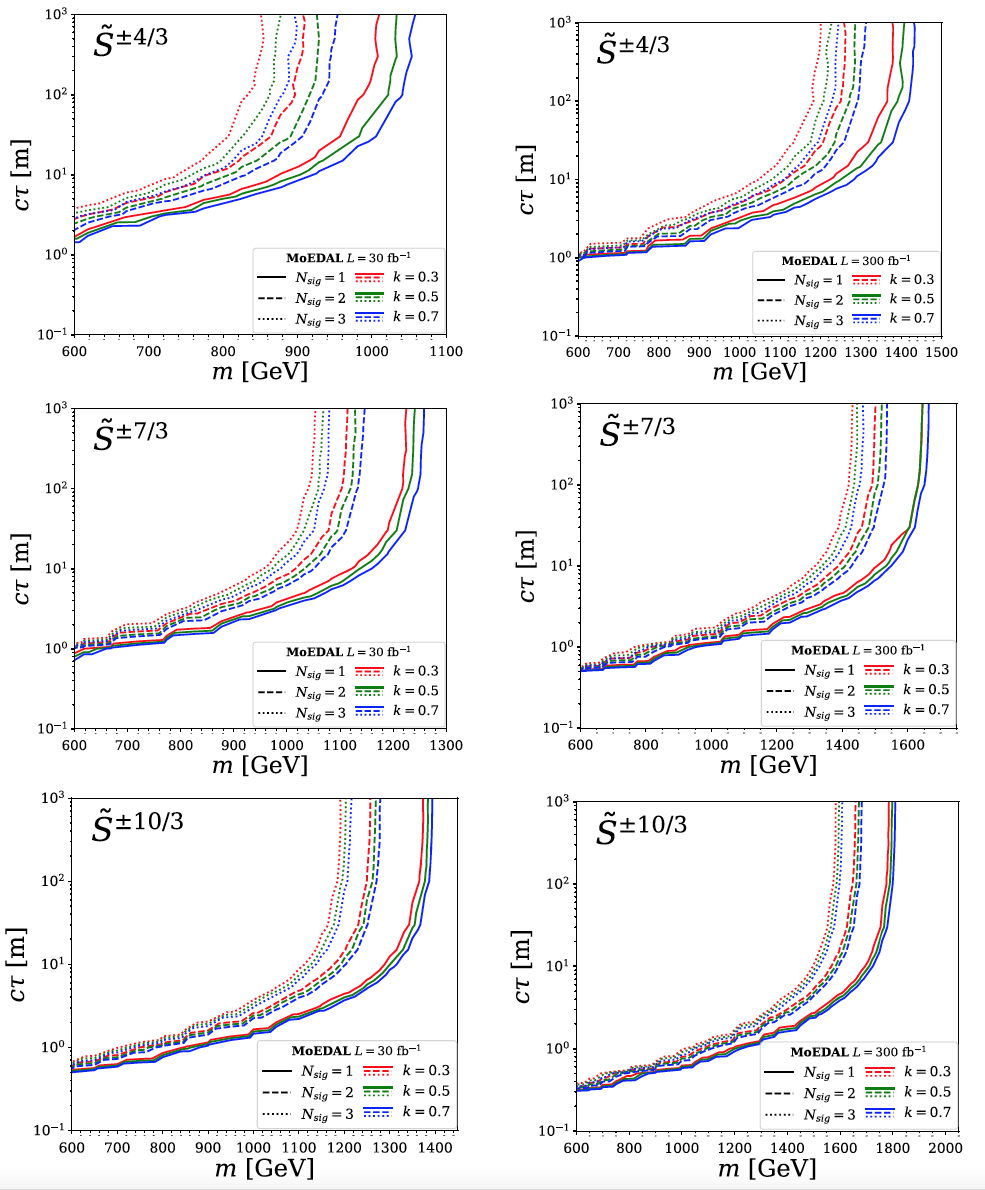}
\caption{The model-independent detection reach at MoEDAL for the LLP particles of neutrino mass Model-2. The contours corresponding to the 
three hadronization parameters of the model, $k = 0.3, 0.5 \text{ and } 0.7$, can be  identified  by the colours  red, green and blue, respectively. The solid, dashed and dotted contours correspond to the detection of 1, 2 and 3 signal events, respectively.
The left(right) panels correspond to an assumed integrated luminosity of  30(300)~fb$^{-1}$~\cite{Hirsch-2021}.}
\label{fig:LLP-coloured}
\end{figure}

 MoEDAL's  expected model independent sensitivity for the  detection of  multi-charged long-lived particles within the framework of Model-2 is shown in Figure~\ref{fig:LLP-coloured}.  As can be seen   the effect of the hadronization parameters on
the detection reach is not strong.

\noindent
\emph{Model Independent Search for Multiply Charged Particle with Charge as high as 8$e$.} We consider multicharged LLPs  produced  in an  open-channel mode, in which the produced  BSM particle-antiparticle pair is  propagates through the detector. In this approach  we consider four types of electrically charged LLPs with spin-0 and spin-1/2 that are colour singlets and triplets \cite{Altakach-2022}. The gauge invariant Lagrangian for the model includes the quantum fields $\phi$ (scalar)  and $\psi$ (fermion). These fields are assumed to be $SU(2)_{L}$ singlets and have the hypercharge $Y=Q/e$, where $Q$ in expressed in units of the electron charge, $e$. 

\begin{figure}[hbt]
\centering\includegraphics[width=0.75\linewidth]{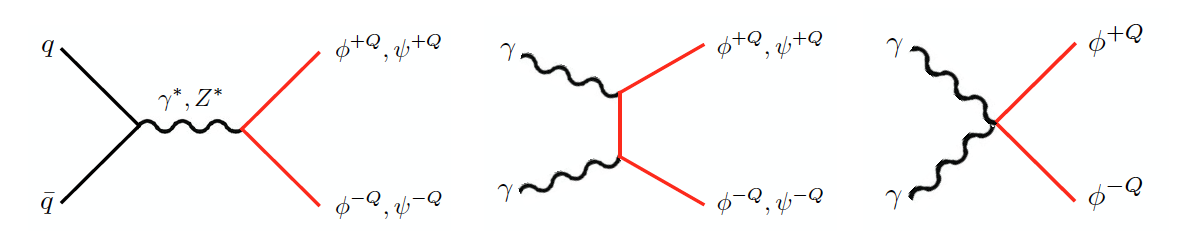}
\caption{The Feynman diagrams for the colour-singlet production modes considered.}
\label{fig:colour-singlets}
\end{figure}

The processes possible for colour singlet particles are shown in Figure~\ref{fig:colour-singlets}. These diagrams include: DY production, photon fusion $t$-channel); and  photon-fusion 4-point interaction for scalars.
The following processes contribute in the case of colour-triplet particles: DY production from a quark-antiquark initial state with $gluon/photon/Z$ exchange; gluon-fusion; gluon-photon-fusion with a mixed QED and QCD $t$-channel interaction; photon-fusion with a $t$-channel interaction and a 4-point interaction for scalars particles~\cite{Altakach-2022}.

\noindent
In addition to open-channel production mode discussed above there is also the possibility  of a ``positronium-like'' bound state composed of BSM particle-antiparticle pairs, that decays to two photons~\cite{Altakach-2022}. This  closed-production mode process cannot be detected by MoEDAL.

\noindent
In the case of ATLAS the  most general and  stringent constraints placed  on the open production mode are from  searches, signified by anomalously high energy deposition in the inner tracker. ATLAS and CMS have performed several such searches for open channel production of  charged LLPs. The  most recent ATLAS results were obtained with an integrated luminosity of 36.1 fb$^{-1}$  at an $E_{\rm CM}$ of  13~TeV \cite{Aboud-LLP-2019}. The latest CMS analysis in this arena \cite{Khachatryan-2016}  was also performed at an $E_{cm}$ of 13~TeV,   using a somewhat  lower integrated luminosity 2.5 fb$^{-1}$. 

\noindent
For particles with a double electric charge or more the above  analyses were performed considering only DY  production and assuming  only colour singlet fermions were involved. In order to recast the ATLAS and CMS  results with the more complete approach sketched above the authors of Ref.~\cite{Altakach-2022} utilized the work of Ref.~\cite{Jager-2019}, extending it by including colourless scalars. They then estimated the projected sensitivity of the ``enhanced'' ATLAS CMS analyses for LHC's Run-3 and HL-LHC.

 \begin{figure}[hbt]
\centering\includegraphics[width=0.9\linewidth]{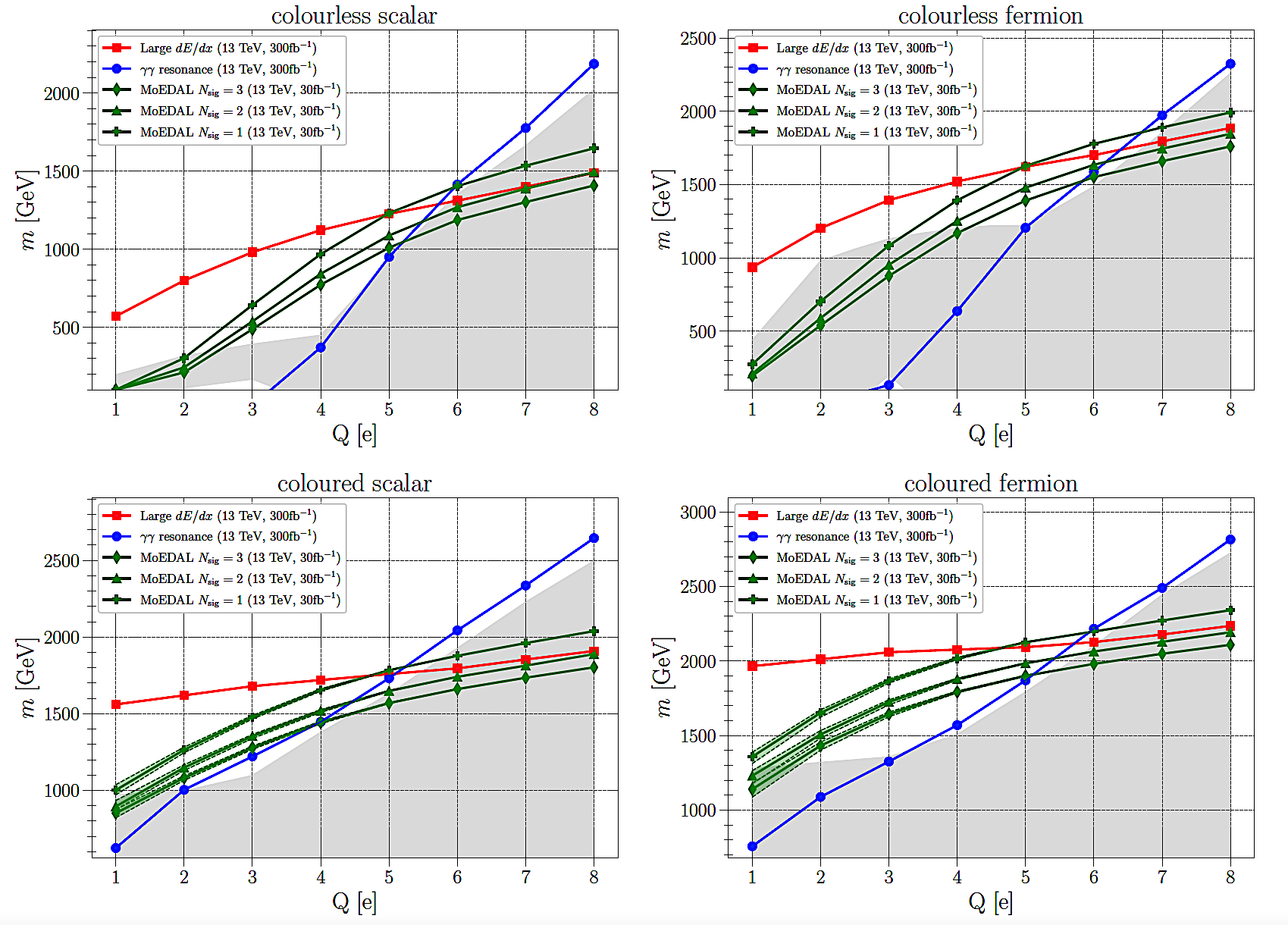}
\caption{Envisaged lower mass limits on new particles: colourless scalars (top left), colourless fermions (top right), coloured scalars (bottom left) and coloured fermions (bottom right)~\cite{Altakach-2022}.
 The integrated luminosity for used for ATLAS/CMS estimates was $L = 3000$~fb$^{-1}$  with  $L = 300$~fb$^{-1}$ for MoEDAL.}
\label{fig:LLP-scalars-fermions}
\end{figure}

\noindent
Figure~\ref{fig:LLP-scalars-fermions} depicts  the estimated  mass limits for Run 3. The green curves represents the expected sensitivity of the MoEDAL experiment, corresponding to 1, 2 and 3 signal events detected. The red and blue curves correspond
to the 95\% CL limits obtained from the recast ATLAS  large dE=dx and diphoton resonance searches, respectively.
As can be seen from Figure~\ref{fig:LLP-scalars-fermions}  MoEDAL is competitive at Run-3. However, for HL-LHC MoEDAL has the potential to give superior results at intermediate charges.  In either case MoEDAL's  totally different systematics and lack of physics backgrounds compared to ATLAS and CMS make it a valuable addition to the assessment of any discovery of multiply charged particles in this arena.

\section{The Phase-1 MAPP Upgrade for MoEDAL}
\begin{figure}[hbt]
\centering\includegraphics[width=0.7\linewidth]{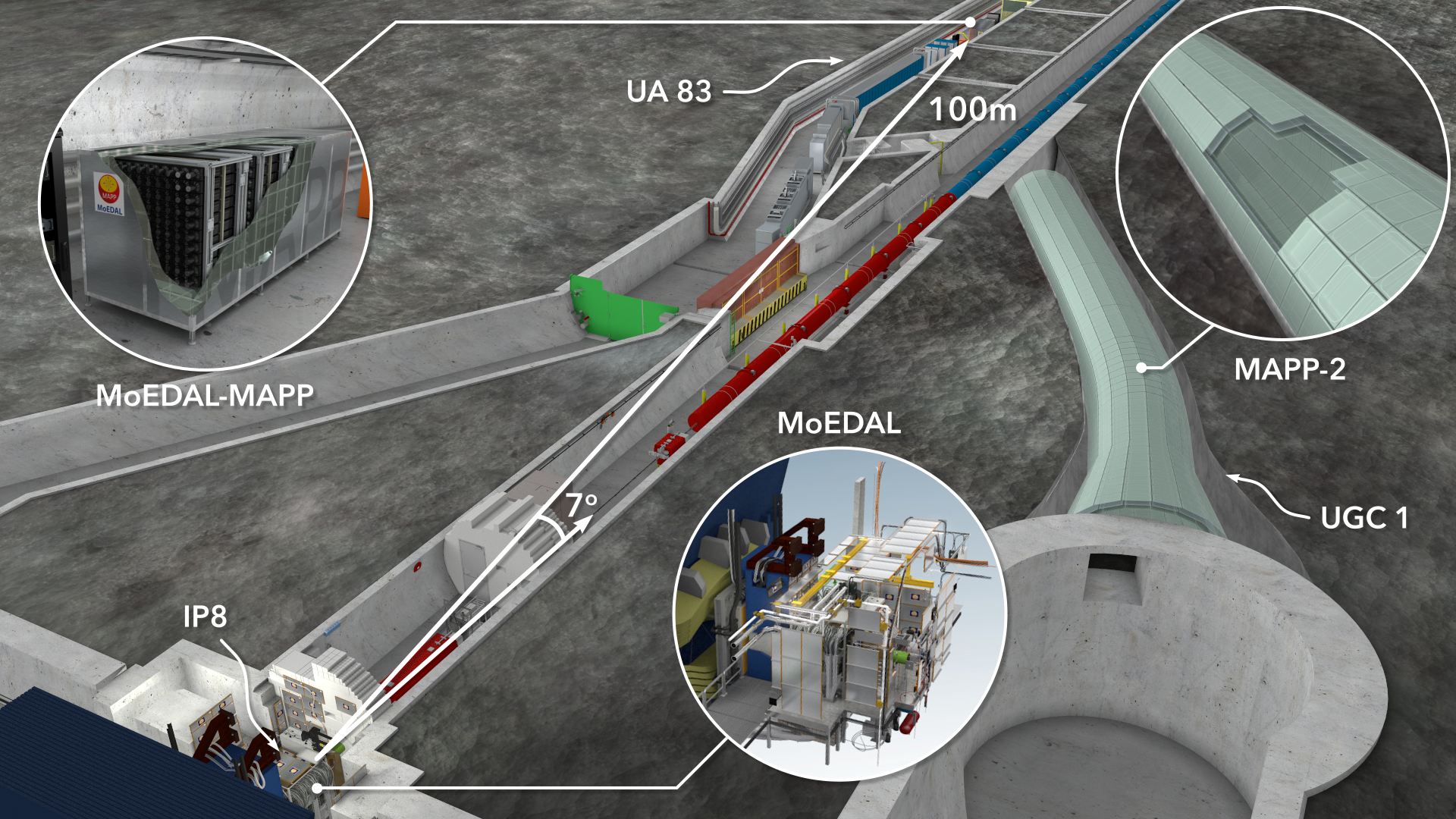}
\caption{ The MoEDAL-MAPP deployment for Run-3 and beyond. MAPP-2 shown on the right is planned for HL-LHC}
\label{fig:MoEDAL-MAPP}
\end{figure}

\noindent
The key part of the MoEDAL Experiment's Run-3 Phase-1 upgrade program at the LHC is the addition to the baseline MoEDAL detector of a new sub-detector called MAPP (MoEDAL Apparatus for Penetrating Particles)~\cite{MAPP-TP}. The MAPP detector is currently being installed in the UA83 gallery adjacent to IP8 on the LHC ring, to take data during Run-3. The MAPP expands the physics reach of MoEDAL to include sensitivity to feebly-charged particles with charge, or effective charge, as low as 10$^{-3}e$. Also, the MAPP detector In conjunction with MoEDAL's trapping detector gives us a unique sensitivity to extremely LLCPs. MAPP also has some sensitivity to very Long-Lived neutral Particles. A picture of the Run3 MoEDAL-MAPP deployment is shown in Figure~\ref{fig:MoEDAL-MAPP}. 

\subsection{The Phase-1 MAPP Detector}
\noindent
In February 2020  the Phase-1 MAPP detector was first presented  to the LHCC as a MoEDAL upgrade for Run-3, as part of a broader upgrade plan encompassing Run-3 and HL-LHC. At this time the
deployment of the Phase-1 MAPP detector was planned for the UGC1 gallery, at a position 55~m from IP8 
at an angle of   6.5$^{\circ}$  to the beam axis. At this position the MAPP-mQP detector would be  
protected by roughly 25~m of rock and by a 110~m rock overburden from SM and cosmic-ray backgrounds. A prototype detector was deployed in UGC1 in 2018  to assess cosmic ray 
and beam related backgrounds.

However, the UGC1 gallery required approximately six months of civil engineering work  to be  commissioned
 as an experimental area, hence there was not sufficient time to complete this work before the start of Run-3. An alternate location for the MAPP-mQP in the UA83 tunnel  was found in June 2021. This new position is already a full part of the LHC infrastructure  with  direct elevator access, with the required access interlocks and  safety measures in place.  Approval for the use of UA83 was  obtained  from the LHC Machine Committee in November 2021. The MAPP detector was approved for installation by the CERN Research Board in December 2021.  
 
 \begin{figure}[hbt]
\centering\includegraphics[width=0.7\linewidth]{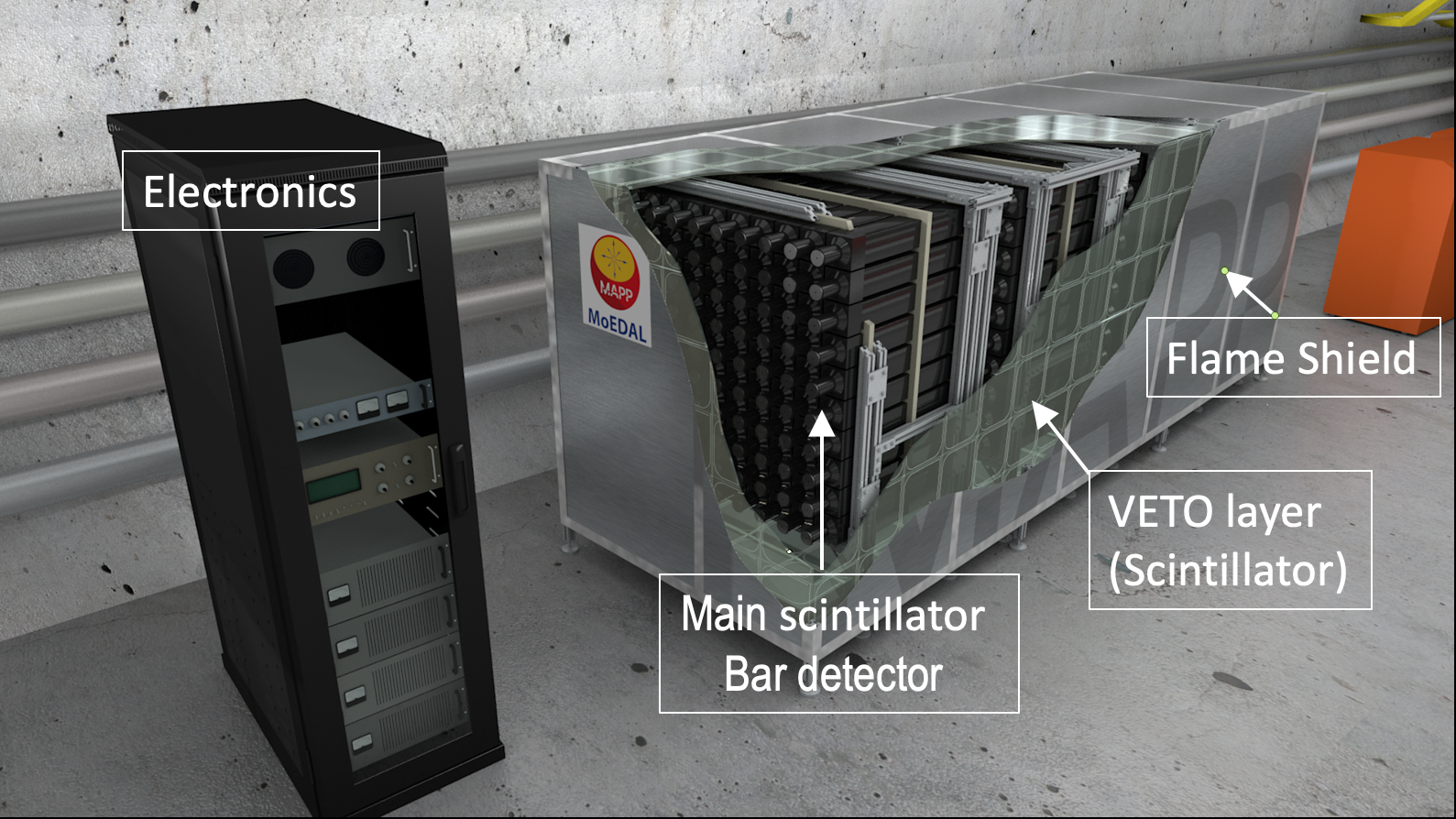}
\caption{The MAPP detector deployed in the UA83 tunnel.}
\label{fig:MAPP-mechanics}
\end{figure}

\begin{figure}[hbt]
\centering\includegraphics[width=0.7\linewidth,angle=-90]{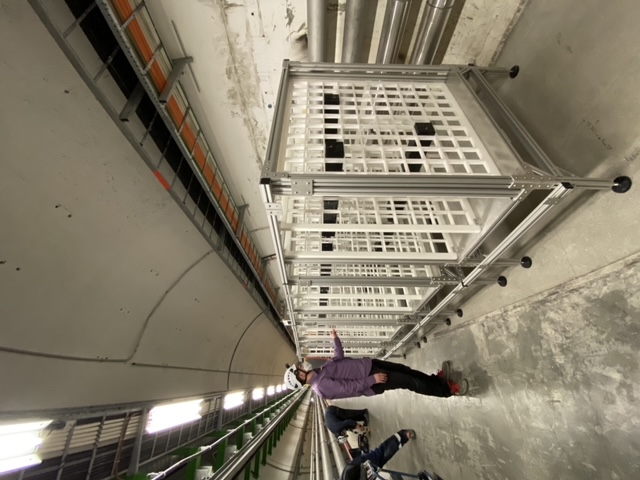}
\caption{ Initial stages of the installation of the MAPP detector in the UA83 tunnel.}
\label{fig:MAPP-installation}
\end{figure}

The MAPP detector~\cite{MAPP-TP} is divided into four equal section as can be seen in  Figure~\ref{fig:MAPP-mechanics}. Each section corresponds to a $10\times10$ array of 100 scintillator each of size 10~cm $\times$ 10~cm $\times$  75~cm  each readout by a low noise 3-inch PMT.  The main support structure is comprised of generic T-bar extruded aluminium construction bars. Each leg of the support structure is adjustable, in order that the  height of the structure can be adjusted to ensure the detector is pointing to IP8. The  scintillator bars in each of the four sections of the MoEDAL-mQP detector are supported by three  matrices machined out of   high-density polyethylene (HDPE) plate. The scintillator bar sections constitute the  main detector which is hermetically encased in a scintillator veto layer. The MAPP detector and VETO layer are completely enclosed in an aluminium  flame shield. The size of the shield is roughly 1.3~m $\times$ 1.5~m $\times$ 4~m. 
Installation of the MoEDAL detector started in December 2021. A photograph of the initial stages of installation is shown in Figure~\ref{fig:MAPP-installation}.

\subsubsection{The MAPP Detector Outrigger} 
\begin{figure}[hbt]
\centering\includegraphics[width=0.7\linewidth]{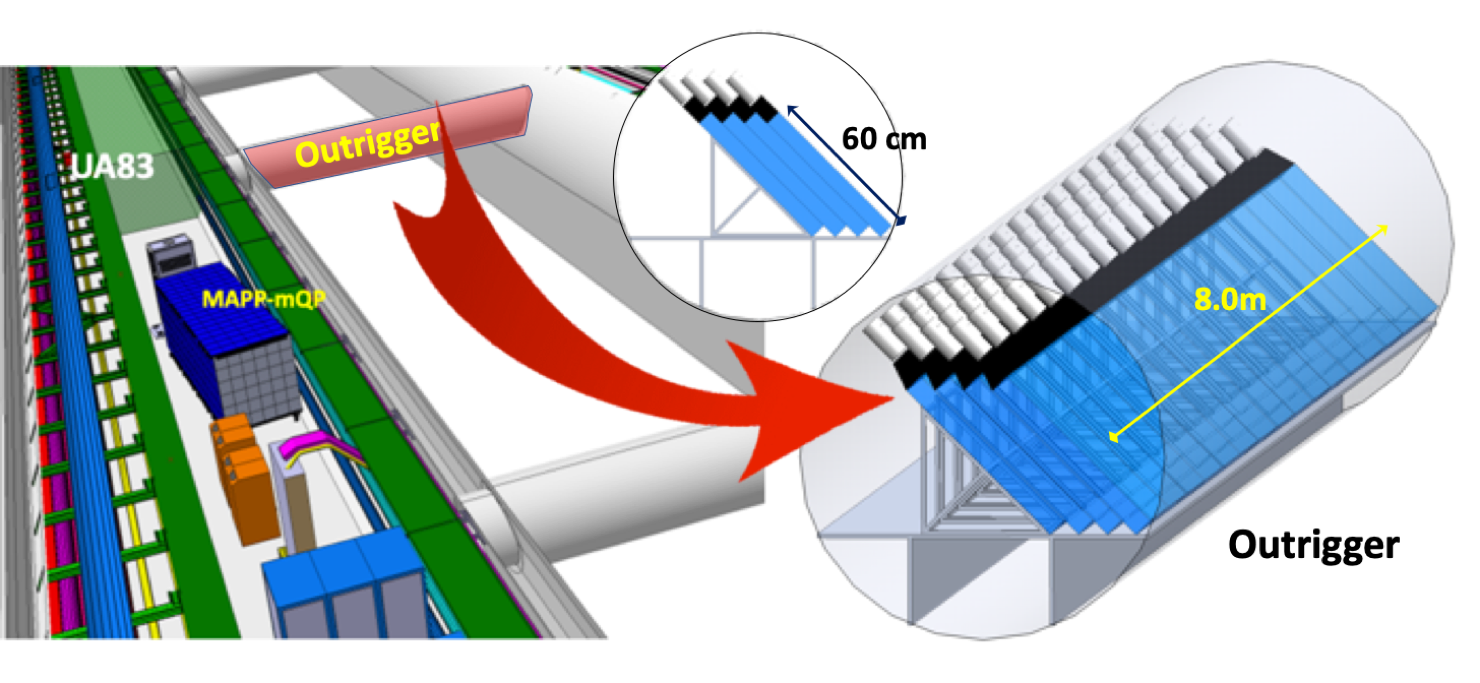}
\caption{A sketch of the outrigger detector for the MAPP detector.}
\label{fig:MAPP-outrigger}
\end{figure}

\noindent
The MoEDAL Collaboration is planning to submit an addendum to the MAPP Technical Proposal to
install an outrigger detector  to improve the acceptance of the  Phase-1 MAPP detector for higher mass
milli-charged particles.  The MAPP-mQP detector is planned for deployment in a circular passage joining the UA83 gallery and the beam tunnel as shown in Figure~\ref{fig:MAPP-outrigger} traversing an angle from roughly 2$^{\circ}$ to 6$^{\circ}$ from the beam axis. It consists of 4 $\times$ 800~cm $\times$ 60~cm $\times$ 5~cm scintillator  planks that each have a cross-sectional area of 4.8 m$^{2}$. Each plank consists of 16 $\times$ 50~cm $\times$ 60 scintillator plates that are each readout by a 2-inch 12-stage PMT. The plates are slanted at 45$^{\circ}$  to give a greater path length in the scintillator for milli-charged particles from IP8 traversing the Outrigger.

\noindent
The increased area of the detector provides greater acceptance for milli-charged particles albeit with a a higher threshold for detection, $\sim0.01e$, than the main MAPP-mQP detector which is $\sim0.001e$. The passage of the milli-charged particle detector through the Outrigger would be marked by a four-fold coincidence between the PMTs in each layer. The milli-charged particle would in this case travel though roughly 28~cm of scintillator rather than the 3~m of scintillator in the case of the MAPP scintillator bar detector. 

\subsubsection{ The MoEDAL-MAPP Detector for Charged LLPs} 
\begin{figure}[hbt]
\centering\includegraphics[width=0.7\linewidth]{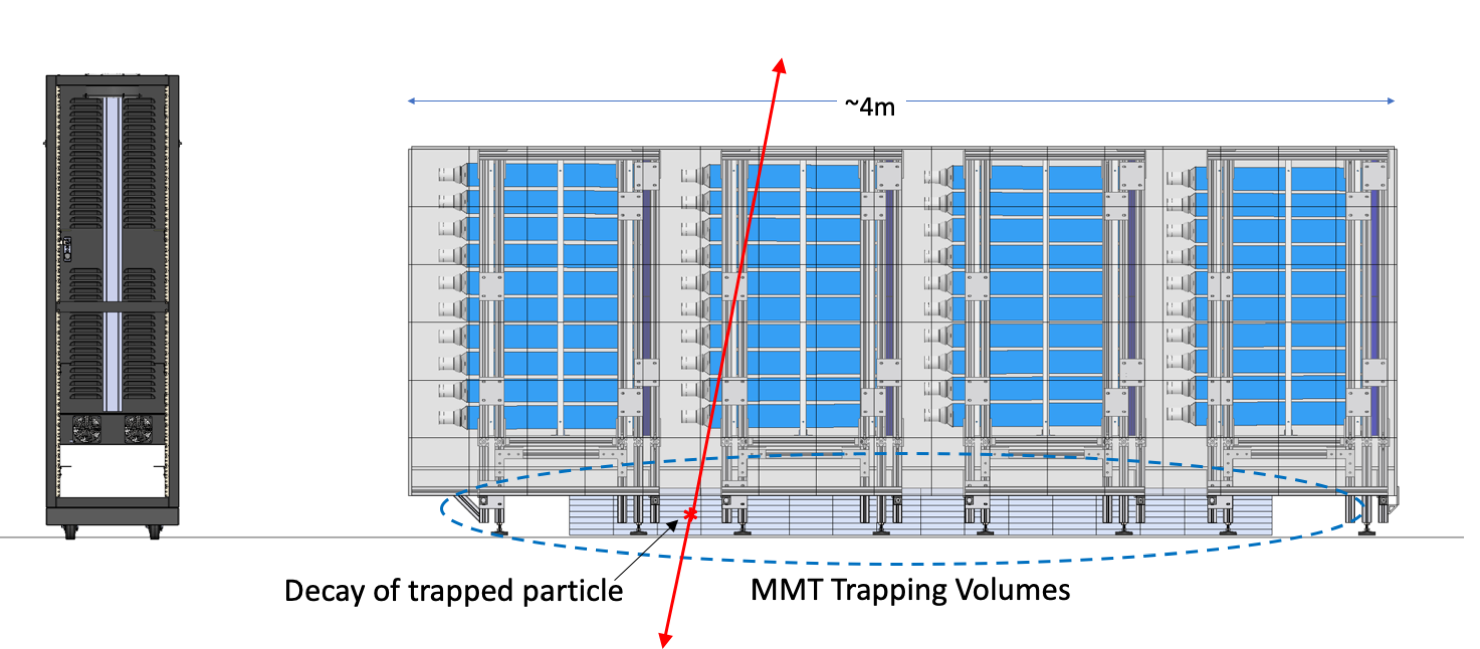}
\caption{A sketch of the MoEDAL-MAPP LLCP detector.}
\label{fig:MAPP-LLCP-detector}
\end{figure}

\noindent
In our initial LoI we described  the MoEDAL Apparatus for very Long Lived Charged Particles (MALL)  detector who purpose is to search for LLCPs.  Our studies have shown that  by placing MoEDAL's 2560 exposed trapping detector volumes,  each of size 2.5~cm $\times$ 2.5~cm $\times$ 19~cm, underneath the MAPP-mQP detector we can monitor them for the decays of  electrically charged particles that have stopped and been captured, obviating the need for the MALL detector. The arrangement of trapping volumes with respect to the MoEDAL-MAPP detector is shown in Figure~\ref{fig:MAPP-LLCP-detector}. This development  will also  be presented in detail in a future addendum to the Technical Proposal. 

\subsection{Physics Reach of the Phase-1 MAPP Detector}
\noindent
The main physics arena of the MAPP detector is that of feebly interacting particles. FIPs are usually defined to be any new, massive or massless,  particle that have extremely small couplings to  SM particles. The size  of these  couplings is possibly due to either to the presence of an slightly broken approximate symmetry  and/or  a large mass hierarchy between particles. Such FIPs are neutral under the SM gauge interactions, although small weak neutral   charge is a possibility. 

\noindent
Searches for BSM  particles have generally concentrated  on weak-scale masses, arising typically from theoretical scenarios such as supersymmetry, and various low-energy realizations  of string theory. However, currently, more attention is  being  placed on lighter  mass   scales than the electroweak one, which  can give rise to FIPs with mass of  the order of GeV$/c^2$ or less  \cite{Alexander:2016,Battaglieri:2017,Beacham:2020,Mitsou:2021tti,Lanfranchi:2011}. Such   light FIPs are  well-motivated dark matter (DM) candidates.  Additionally, they provide a possible  explanation for:  the strong  CP problem,  the origin of neutrino masses;  and,  the  universe's baryon asymmetry.  

\noindent
Indirect interactions of FIPs with SM particles are possible through ``portals'' \cite{Beacham:2020,Batell-2009}. The class of  particles defined this way are seen as being part of a dark, or hidden, sector. The ``portal framework''  is a  formal  way to describe all  interactions between a dark sector and the SM.
This is achieved by constructing the interaction Lagrangian: 
\begin{equation} 
\mathcal{L}_{\text{total}} = \mathcal{L}_{\text{SM}} + \mathcal{L}_{\text{DS}} + \mathcal{L}_{\text{portal}}, \quad \text{with }\quad \mathcal{L}_{\text{portal}} =  \sum(O_{\text{SM}} \times O_{\text{DS}}),
\end{equation}
where $O_{\text{SM}}$  is an operator composed of SM fields and $O_{\text{DS}}$ an operator composed from the dark sector fields.  The sum is over a number of possible operators  of different dimension and compensation. To guarantee the SM neutrality of the dark sector both these operators must be gauge singlets. This requirement places a stringent restriction on the allowed types of interaction.

\noindent
The portals with the  lowest dimension include: the scalar portal -  mediated by a new scalar mixing to the SM Higgs boson;
the vector portal, mediated by a dark gauge boson characterized as a  dark photon; 
the fermion portal, mediated by a heavy neutral lepton interacting with the Higgs boson and one
of the left-handed SM doublets; the pseudoscalar portal, mediated by an axion, or an axion-like particle (ALP) that  couples
 to fermion and gauge fields.  The axion portal  is suppressed by the axion decay constant. However, the scalar, vector and fermion 
portals are renormalizable and not suppressed by any physics scale, which  can potentially be very large.

\noindent
It turns out that FIPs  are currently  only weakly constrained by current direct experimental observations.
Currently, the  majority of FIP related work involves milli-charged particle, heavy neutral leptons 
 (HNLs),  axion-like particles (ALPs),  dark photons and dark Higgs bosons.    A common scenario is where one  considers a  mCP  coupled through a very light  kinetically mixed dark photon. Although the mCP does not carry SM electroweak quantum  numbers it  behaves  as a particle with a tiny electric charge. 
 
 \noindent 
  The Phase-1 MAPP detector is primarily designed to  the search for milli-charged particles with charge as low
 as $10^{-3}e$. The detector is competitive with the milliQan detector \cite{milliQan}  that will also be deployed for Run-3 and covers a different pseudo-rapidity range. In the event of the discovery of a milli-charged particle by MAPP-mQP and milliQan, a signal seen in two different detectors with their different systematics would provide the necessary confirmation of a discovery.
 
 \noindent
 Additionally, the MoEDAL and MAPP-mQP detectors working together will provide a superior sensitivity for LLCPs as well as the unprecedented ability, at a collider, to detect and measure lifetimes that extend out to 10 years or more. The volume of the MAPP-mQP detector and veto system can also be used to search for LLPs. Importantly, the detector covers a different pseudo-rapidity region than FASER and in that sense provides a complementary coverage. Additionally, MAPP-mQP's limited but useful sensitivity in this arena could in some circumstances provide a confirmation of a signal observed by FASER \cite{FASER}.

\subsubsection{Searching for mCPs with the Phase-1 MAPP Detector}
\noindent
Fractionally charged particles with charges well below approximately $e$/3 of the  LHC experiments cannot be detected by the general purpose  LHC experiments, ATLAS and CMS. This is mainly due to the fact the detectable ionization of the particle is proportional to the square of the charge. Thus, the search for such mCPs  falls to dedicated search experiments such as MoEDAL-MAPP and milliQan. In the SM, the only elementary particles with fractional electric  charges are quarks that  cannot be observed  due to colour confinement. However, there are a numbers of BSM scenarios in which unconfined fractionally charged particles are predicted including  dark sector models and  some Superstring models, for example, Ref.~\cite{Superstring-mCPs}. 

\begin{figure}[hbt]
\centering\includegraphics[width=0.45\linewidth]{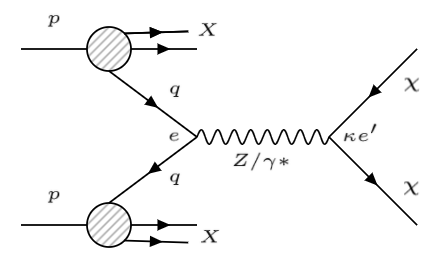}
\caption{A Feynman diagram for the  DY production of mCP-pairs, $\chi\chi$, in $pp$ interactions.}
\label{fig:mQP-Feynman}
\end{figure}

\begin{figure}[hbt]
\centering\includegraphics[width=0.7\linewidth]{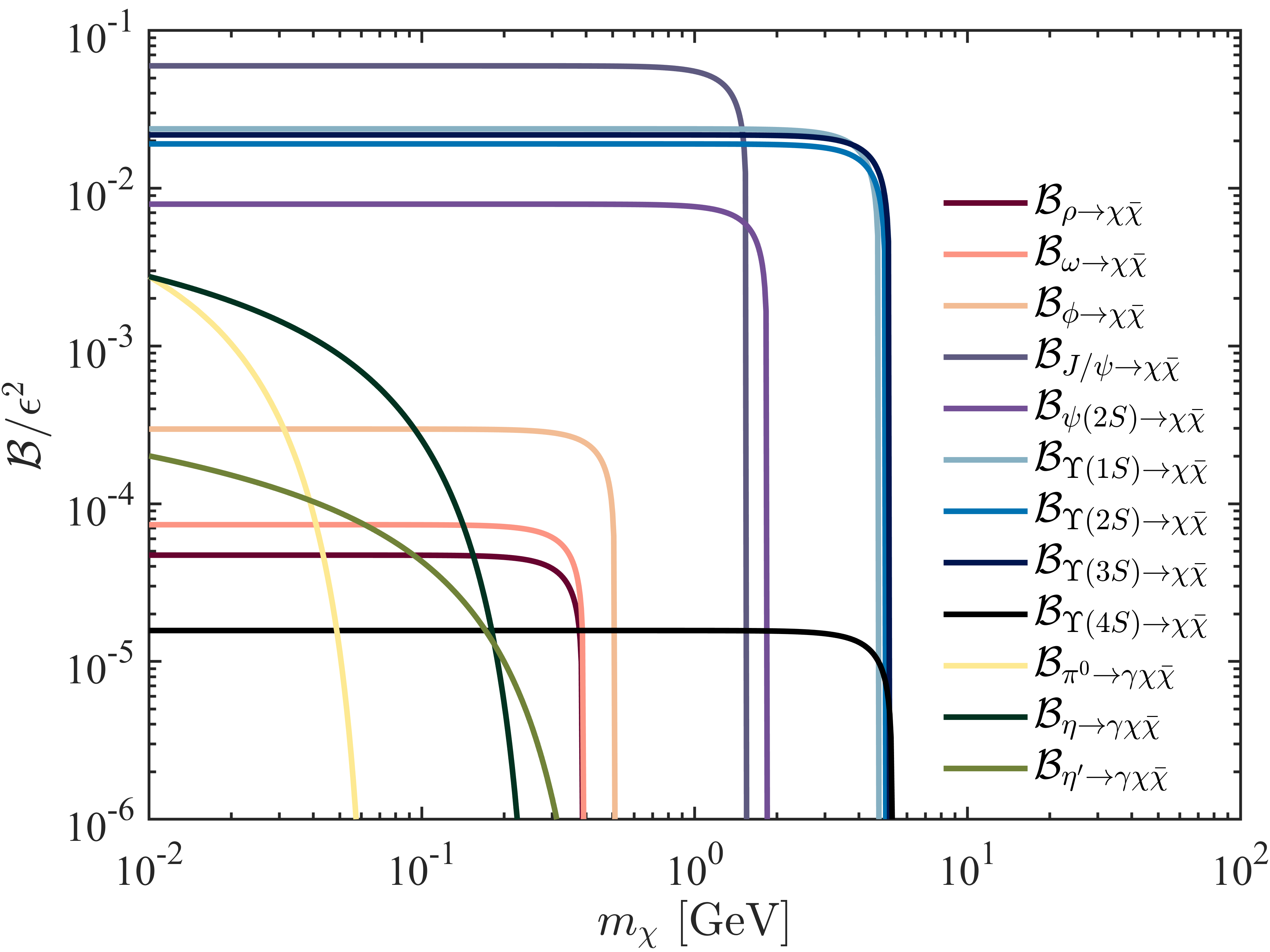}
\caption{Various branching ratios for decays to milli-charged particle pairs.}
\label{fig:BRs}
\end{figure}

\begin{figure}[hbt]
\centering\includegraphics[width=0.8\linewidth]{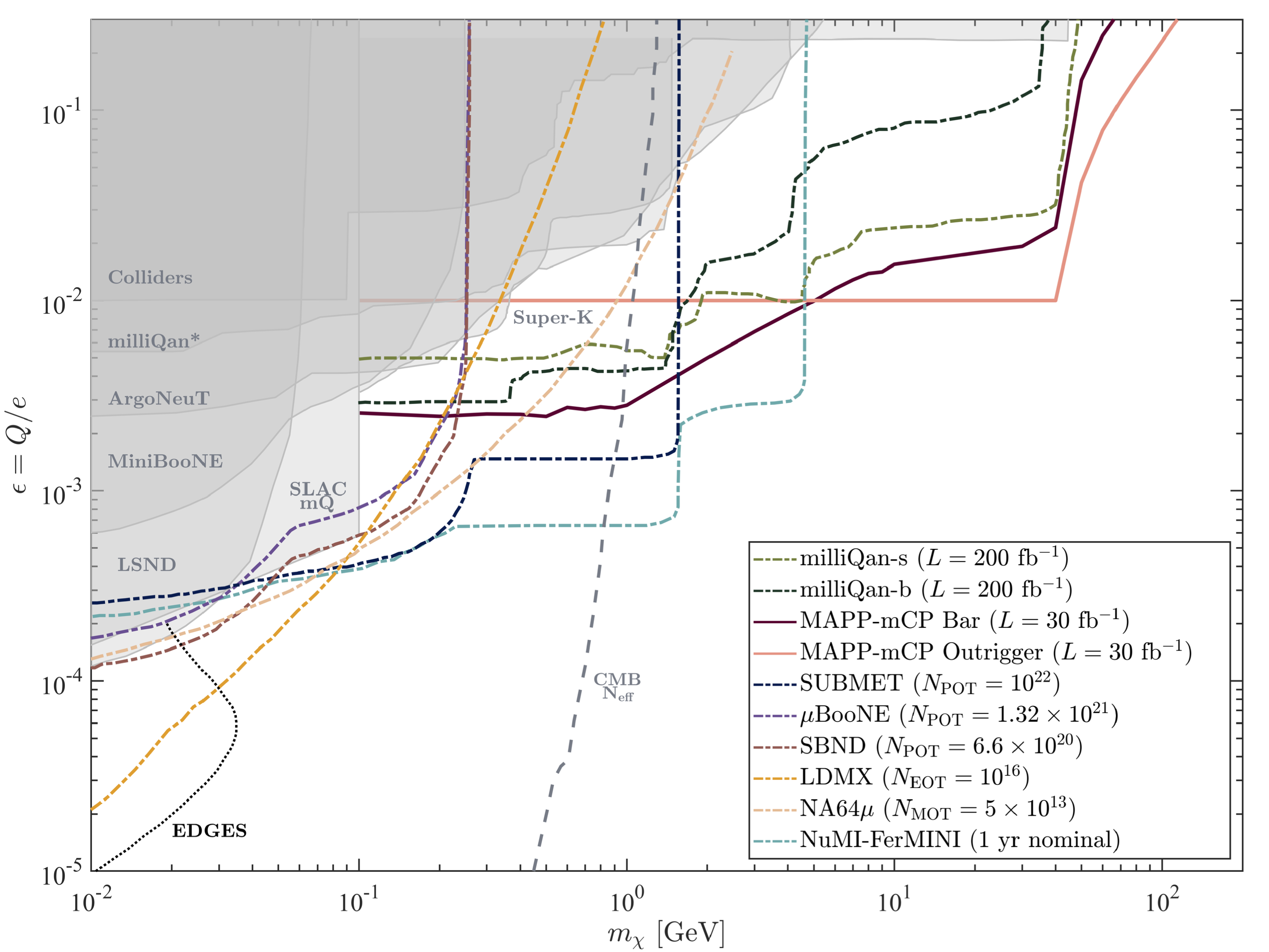}
\caption{Direct bounds from accelerator based searches~\cite{Prinz-1998,Magill-2019,Aguilar-2018,Aguilar1-2018,Davidson-2000, Acciarri-2020,Ball-2020,Plestid-2020,LSND:2001akn} and indirect bounds from the effective number of neutrinos from Planck~\cite{Brust-2013,Vogel:2013raa} (gray regions). The projected sensitivity for mCPs, for models with a massless dark photon are presented for milliQan (for the slab (s) and bar (b) detectors) and MAPP-mQP (for the bar (B) and Outrigger (O) detectors) in $p-p$ collisions at $\sqrt{s}=14$~TeV (colored lines)~\cite{Magill-2019, Ball-2021,Kim-2021,Berlin-2019,Gninenko-2019}.}
\label{fig:MAPP-mQP-limits}
\end{figure}

\begin{figure}[hbt]
\centering\includegraphics[width=0.8\linewidth]{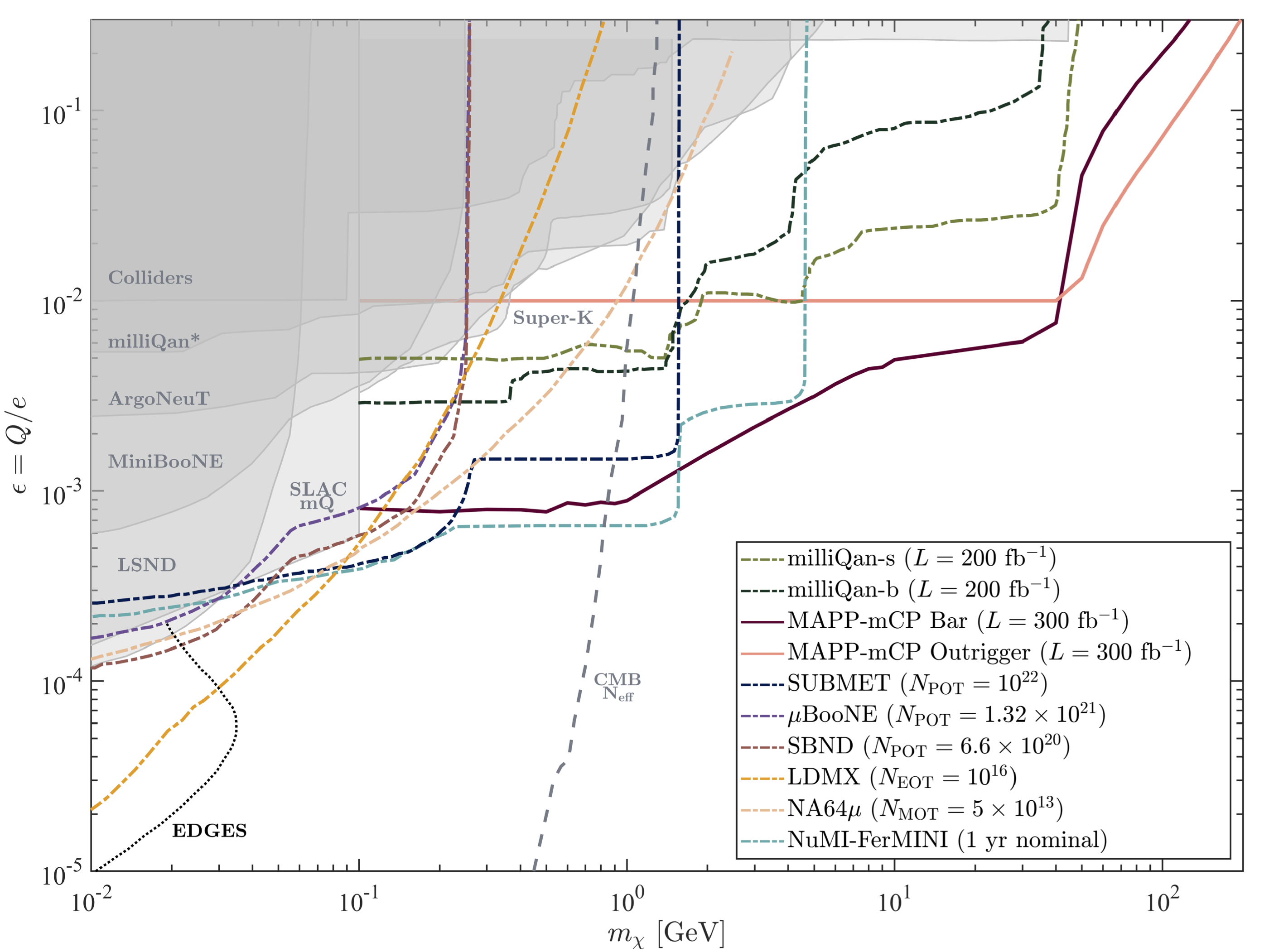}
\caption{The caption is the same as in Figure~\protect\ref{fig:MAPP-mQP-limits} except that the luminosity utilized for the study is that
expected at HL-LHC, namely 300~fb$^{-1}$.}
\label{fig:MAPP-mQP-limits-300}
\end{figure}

\noindent
We consider  here a class of FIPs that has a milli-charge as small as 10$^{-3}e$ or lower. A common scenario is from a Dark Sector model where one considers an mCP coupled through a very light kinetically mixed dark photon \cite{Holdom-1986}. We explore here the possibility  of mCPs arising  from a   model-independent  scenario \cite{Haas-2015}  in which a new massless (Abelian) $U'(1)$ gauge field, the dark photon $A'_\mu$, couples to the SM hypercharge gauge field, $B_{\mu\nu}$, through a mixing term $-(\kappa/2) A^\prime_{\mu \nu} B^{\mu \nu}$, with $\kappa$ a free parameter. A new massive dark-fermion, $\chi$, with mass $m_\chi$ that couples to the dark photon gauge field $A'_{\mu}$ is also predicted. This particle is charged under this new $U(1)$ field with charge $e'$. 
This mixing term can be eliminated by redefining the dark photon field,
$A'_{\mu} \rightarrow A'_{\mu} + \kappa B_{\mu}$. This field redefinition reveals the coupling of the charged matter field $\chi$ to the SM hypercharge that is clear in the new Lagrangian, 
\begin{equation} 
\mathcal{L} = \mathcal{L}_{\text{SM}} - \frac{1}{4}A'_{\mu\nu}A^{'\mu\nu} + i\overline{\chi}\left( \slashed{\partial}
+ ie' \slashed{A'}   - i\kappa e'\slashed{B} +  im_\chi \right)\chi. 
\end{equation}
The above redefinition makes it plain that in the visible sector the fermionic field $\chi$ acts as milli-charged matter field, with milli-charge $\kappa e'$,  that couples to the photon and the $Z^0$  with a charge $\kappa e' \cos\theta_{W}$ and $\kappa e' \sin\theta_{W}$, respectively. Thus, if we define the fractional charge in terms of the electric charge we have $\epsilon = \kappa e' \cos\theta_{W}/e$. We study the production of  mCPs in $pp$ collisions primarily through  the Drell-Yan mechanism, shown in Figure~\ref{fig:mQP-Feynman}, but also through the channels shown in Figure~\ref{fig:BRs}. The sensitivity of the MAPP-mQP detector deployed at UA83 to milli-charged particles produced via DY only is shown in Figure~\ref{fig:MAPP-mQP-limits} and Figure~\ref{fig:MAPP-mQP-limits-300} for Run-3 the HL-LHC, respectively. Moreover, MAPP-mQP can detect a heavy neutrino with a large enough electric dipole moment, experimentally similar to an mCP, considered to be a member of a fourth generation lepton doublet~\cite{Frank:2019pgk}.

\subsubsection{Searching for LLPs with the Phase-1 MAPP Detector}
\noindent
The Phase-1 MAPP detector is not primarily designed to detect the decays of LLPs. However, the MAPP detector consists of a dense volume of segmented scintillator detectors with deployed volume of 1.2~m $\times$ 1.2~m  $\times$ 4~m surrounded by a hermetic veto layer that we envisage will enable us to search for LLPs. We consider below the fiducial sensitivity of the Phase-1 MAPP detector assuming the the detection efficiency for all LLPs decaying within the fiducial volume of the detector is 100\%.

\begin{figure}[hbt]
\centering\includegraphics[width=0.5\linewidth]{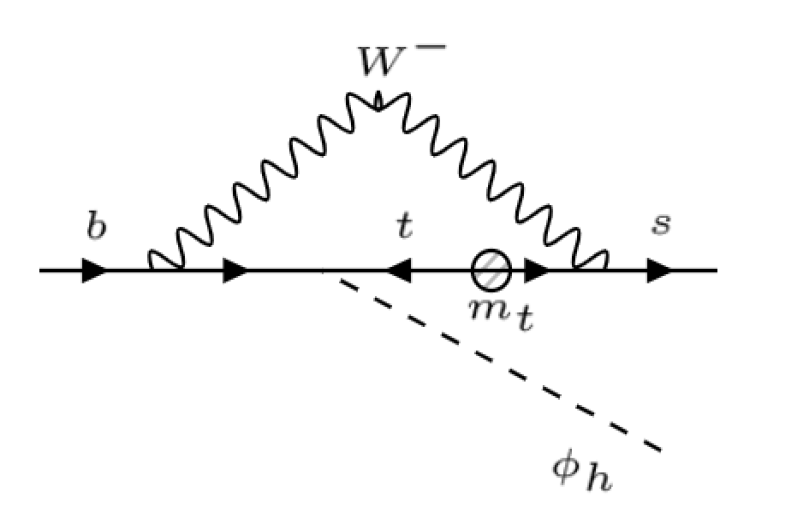}
\caption{Example one-loop Feynman diagram for $b \rightarrow s\phi_{h}$.}
\label{fig:btohiggs}
\end{figure}

\noindent
To illustrate the physics reach of the Phase-1 MAPP detector for  LLP dark Higgs bosons, we used a well studied  benchmark involving the decay of dark Higgs bosons \cite{Feng-2018,Gligorov-2018}. In this physics process a dark Higgs mixing portal allows the exotic inclusive $B$ decays,  $B \rightarrow X_{s}\phi_{h}$, where $\phi_{h}$ is a light CP-even scalar that mixes with the SM Higgs with mixing angle  $\theta$ that is expected  to be much less than one due to existing experimental and theoretical constraints. An example
 of a one-loop diagram that contributes to this decays is shown in Figure~\ref{fig:btohiggs}. 

\noindent
A  simple Lagrangian that includes this new dark Higgs mixing is \cite{Ariga-2019}:
\begin{equation}
\mathcal{L} = \mathcal{L}_{\text{Kin}} + \mathcal{L}_{\text{DS}} + \mu^{2}S^{2} - \frac{\lambda_{S}}{4}S^{4} + \mu^{2}|H|^{2} - 
\lambda|H|^{4} -\epsilon_{h}S^{2}|H|^{2}
\end{equation}
where $H$ is a SM Higgs-like field, $S$ is a real scalar field, $\epsilon_{h}$ is the portal coupling, $\lambda$ and 
$\mu$ are real constants and $\lambda_{S}$ and $\mu_{S}$ are free parameters. The final term is the Higgs portal 
quartic scalar interaction.

\begin{figure}[hbt]
\centering\includegraphics[width=0.9\linewidth]{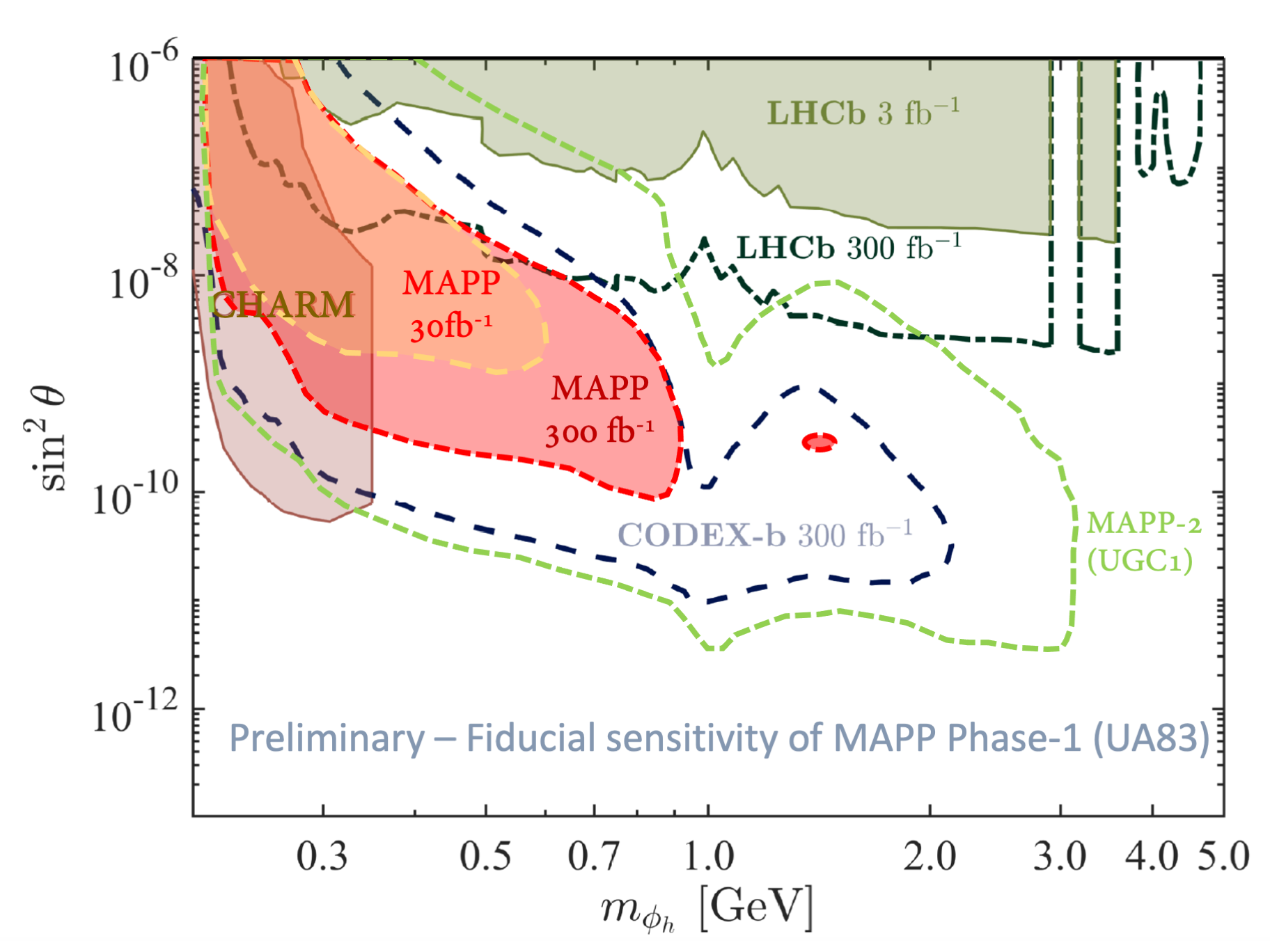}
\caption{Preliminary fiducial sensitivity of MAPP to the dark Higgs model described in the text.}
\label{fig:LLP-MAPP-1}
\end{figure}

\noindent
To explore the fiducial efficiency of MAPP-1 we studied the decay $B \rightarrow K\phi_{h}$, where the $\phi_{h}$  decays to $\mu^{+}\mu^{-}$. The maximum fiducial sensitivity of MAPP-1  for this channel  can be estimated by counting all the decays that occur within the fiducial volume defined by the hermetic VETO wall surrounding the MAPP-1 bars. The resulting maximum MAPP-1  sensitivity to this channel  is shown in Figure~\ref{fig:LLP-MAPP-1} determined by requiring a minimum of 3 decays within the fiducial volume.  Studies utilizing a full simulation of the MAPP-1 detector as well as the surrounding material and cosmic ray background are currently underway. Although, the full simulation will certainly show a reduction of sensitivity compared with that shown in Figure~\ref{fig:LLP-MAPP-1} MAPP-1 will have some useful sensitivity to LLPs during Run-3, while the main dedicated  LLP searches  are planned for Run-4 several years later. The limited sensitivity of MAPP-1 to LLPs, even at HL-LHC (300~fb$^{-1}$) will be enhanced by MAPP-2 that is planned for installation during the long shutdown (LS3) prior to the startup of HL-LHC.  MAPP-2 is the topic addressed in the next section.

\section{The Phase-2 MAPP-2 Upgrade to MoEDAL}
\begin{figure}[hbt]
\centering\includegraphics[width=1.0\linewidth]{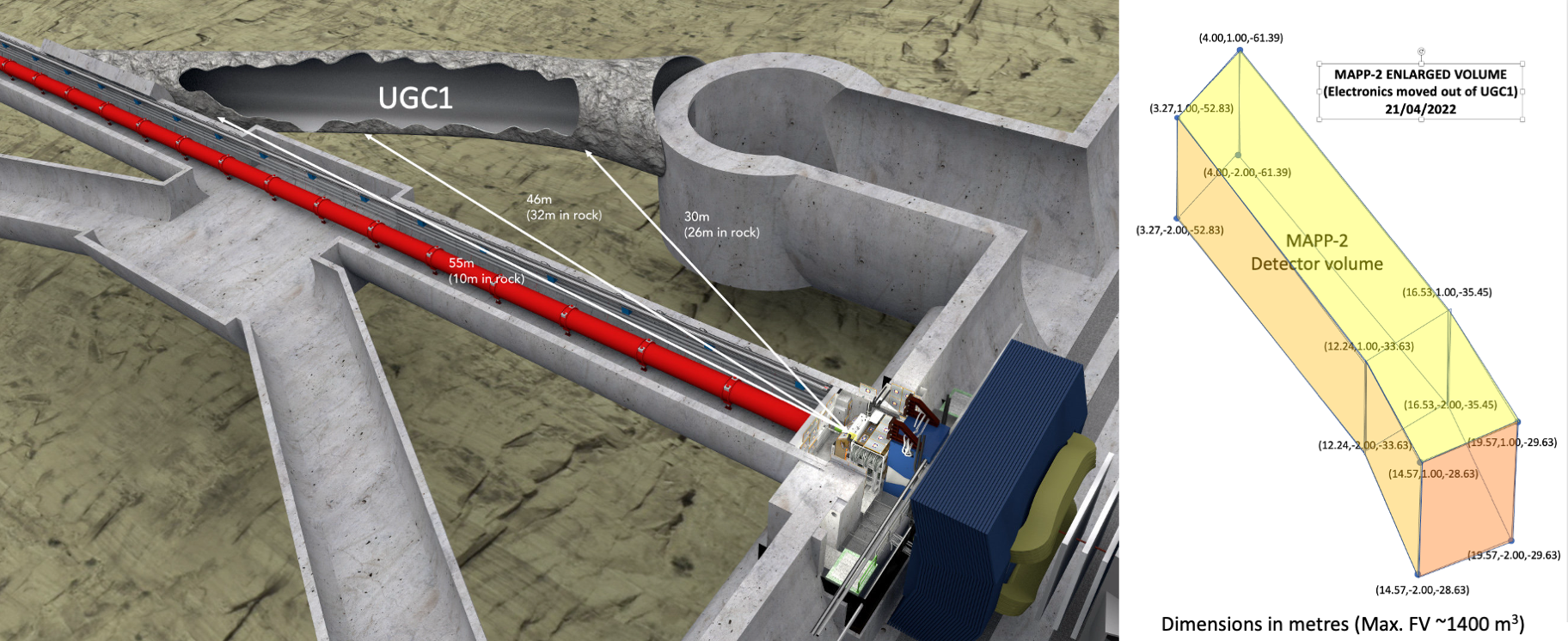}
\caption{The fiducial decay volume defined by the UGC1 gallery.}
\label{fig:MAPP-2-FV}
\end{figure}

\noindent
In Phase-2 of the twenty year plan for MoEDAL we plan to install MAPP-2 in the UGC1 gallery adjacent to IP8, as
shown in Figure~\ref{fig:MoEDAL-MAPP}. In this case the whole UGC1 gallery will be instrumented as a decay zone for LLP from IP8. The  maximum fiducial volume in this case is shown in Figure~\ref{fig:MAPP-2-FV}. the MoEDAL collaboration carried out a readiness review for the UGC1 gallery in conjunction with LHCb and CERN in 2020. Formerly this gallery was used to house a boring machine used to create the LEP-ring tunnel. It was determined at that time that around 300~kCHF -- 400~kCHF of civil engineering work would be required to bring this gallery up to specification as an experimental area.  MoEDAL's plan is to perform this work in LS3 to enable MAPP-2 to be deployed for HL-LHC.

\begin{figure}[hbt]
\centering\includegraphics[width=0.8\linewidth]{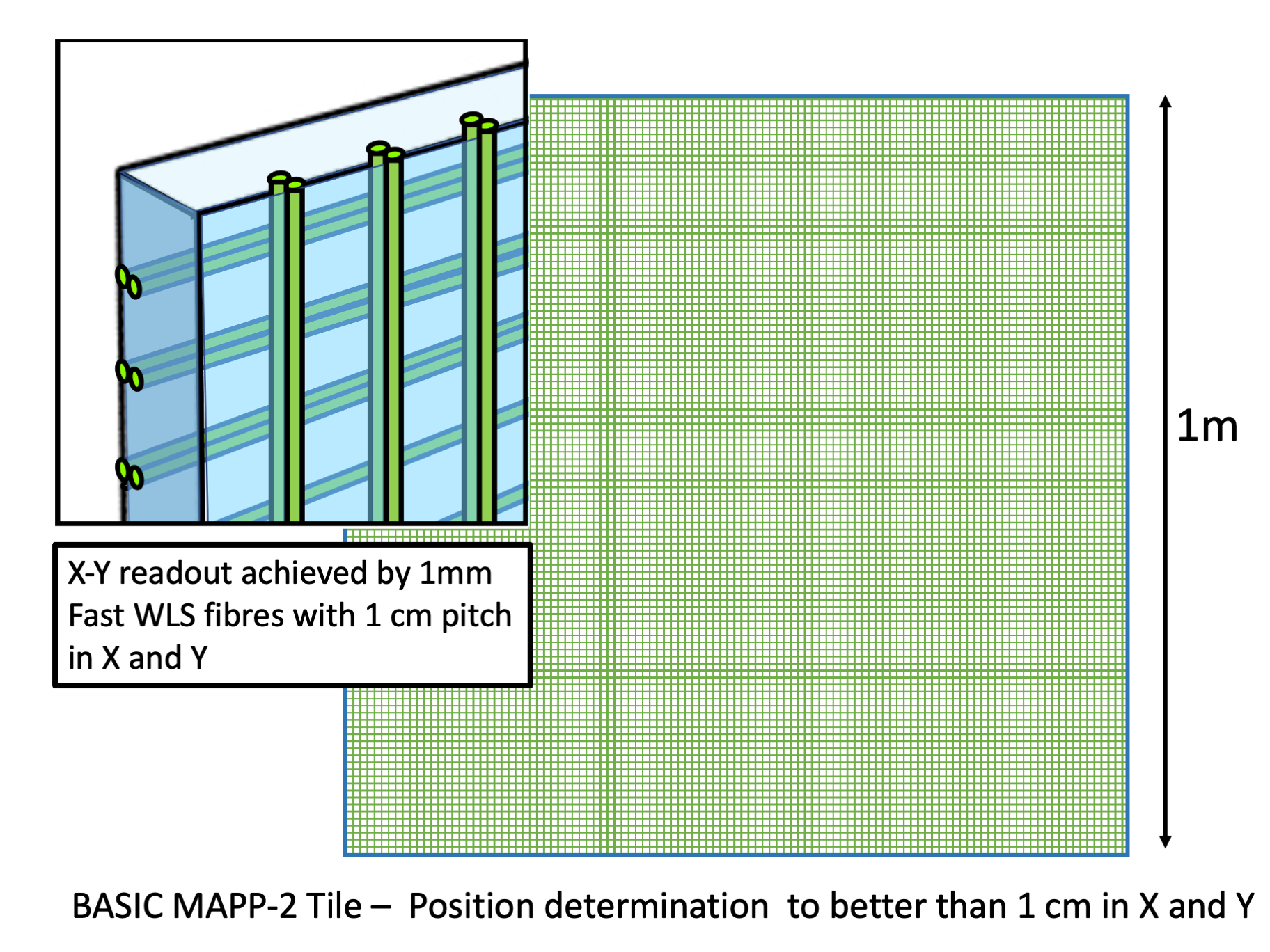}
\caption{A basic  large tile from which  the MAPP-2 detector is constructed.}
\label{fig:X-Y-WLS}
\end{figure}

\noindent
As can be seen from Figure~\ref{fig:MoEDAL-MAPP}, the  end of the UGC1 gallery angle  is some 55~m from 
IP8 and the mouth of the gallery is some 24~m from IP8 spanning an angle to the beam of around 5$^{\circ}$ to 
25$^{\circ}$. The volume enclosed is around 1200~m$^{3}$.  The  MAPP-2  detector design would see the UGC1 gallery 
encased in three layers of scintillating tiles readout by fast WLS fibres attached to silicon photomultipliers (SiPMs). The WLS fibres form an $X-Y$ grid, with the $X$ aligned fibres on one surface of the tile and the $Y$-aligned fibres on the other surface. The pitch of the fibres is 1~cm. This arrangement is illustrated in Figure~\ref{fig:X-Y-WLS}. If all fibres are instrumented the position resolution of a particle traversing the file would be better than 1cm in both the $X$ and $Y$ coordinates. The three layers of scintillators detectors would allow the reconstruction of the trajectory  of tracks from the decay of a LLP. Multiple tracks would allow the reconstruction of  the vertex of the decay. 

\noindent
The UGC1 gallery is protected by 30~m of rock from collision products from the IP8. In order to severely reduce SM particle backgrounds that penetrate this rock protection  a  VETO layer is deployed to  line the face of the UGC1 gallery that faces IP8. Although the UGC1 gallery is protected by cosmic rays by  a  100~m -- 110~m rock overburden, cosmic backgrounds are still an issue in MAPP-2. In order to further protect  against cosmic backgrounds the three layers of scintillator detectors that encase the UGC1 gallery, forming the MAPP-2 detector, are spaced to allow the determination  of whether detected tracks are entering or leaving MAPP-2's fiducial  volume.

\subsection{Physics Reach of the Phase-2 MAPP-2 Detector}
\begin{figure}[hbt]
\centering\includegraphics[width=0.9\linewidth]{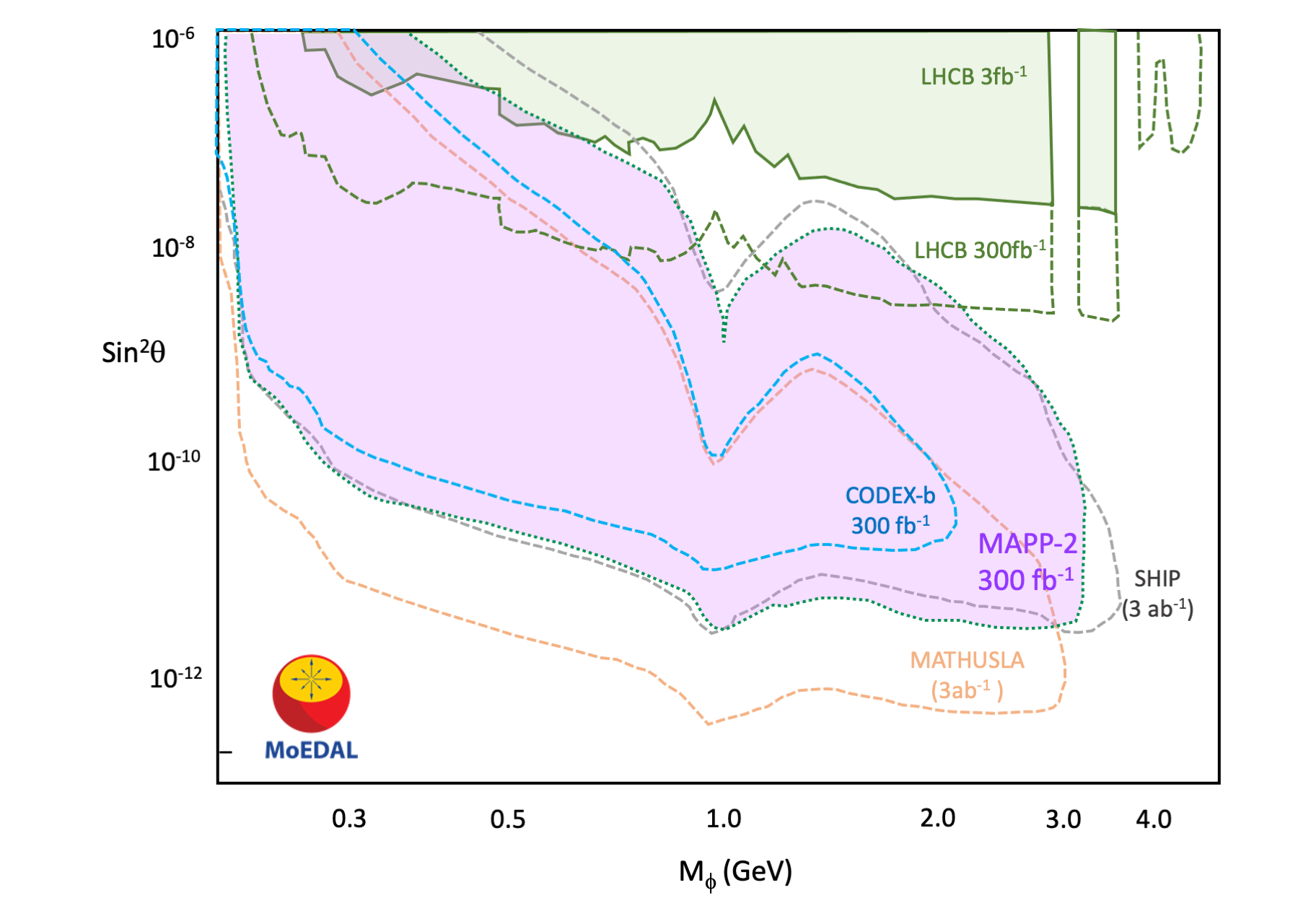}
\caption{MAPP-2 reach for $ B\rightarrow \chi_{s}\phi$  in the $\sin^{2}_\theta$ - $\phi$ plane where
100\% detector/tracking efficiency  is assumed.}
\label{fig:scalar-MAPP2-limits}
\end{figure}

\noindent
The MAPP-2 detector, deployed for the HL-LHC  will enhance the fiducial volume used to detect the decays of LLPs over the MAPP-1 detector now being deployed for Run-3, by a factor of a few hundred times. To illustrate the physics reach of the Phase-2 MAPP detector for  LLPs dark Higgs bosons, we again used the well  studied  benchmark involving the decay of dark Higgs bosons \cite{Feng-2018,Gligorov-2018} that was described previously. In Figure~\ref{fig:scalar-MAPP2-limits} the maximum sensitivity of the MAPP-2 detector is compared with that of the CODEX-b \cite{CODEX-b-limits}, CHARM \cite{CHARM-limits}, LHCb \cite{LHCb-limits,CODEX-b-limits}, MATHUSLA \cite{MATHUSLA-limits,CODEX-b-limits} and  SHiP \cite{SHIP-limits,CODEX-b-limits}. The MAPP-2 design and detector technology choices were made not only to provide competitive functionality but also low cost. We estimate the cost of MAPP-2 to be approximately \$3M (US). 

\noindent
Although, the MAPP-2 contour is expected to shrink slightly when detector efficiency and various background sources are 
considered, it is clear that MAPP-2 is a competitive and complementary detector which in the physics process considered here 
covers the important region between LHCb and MATHUSLA, at the same time covering a significant part of the possible discovery 
contour of MATHUSLA. 

\begin{figure}[hbt]
\centering\includegraphics[width=0.8\linewidth]{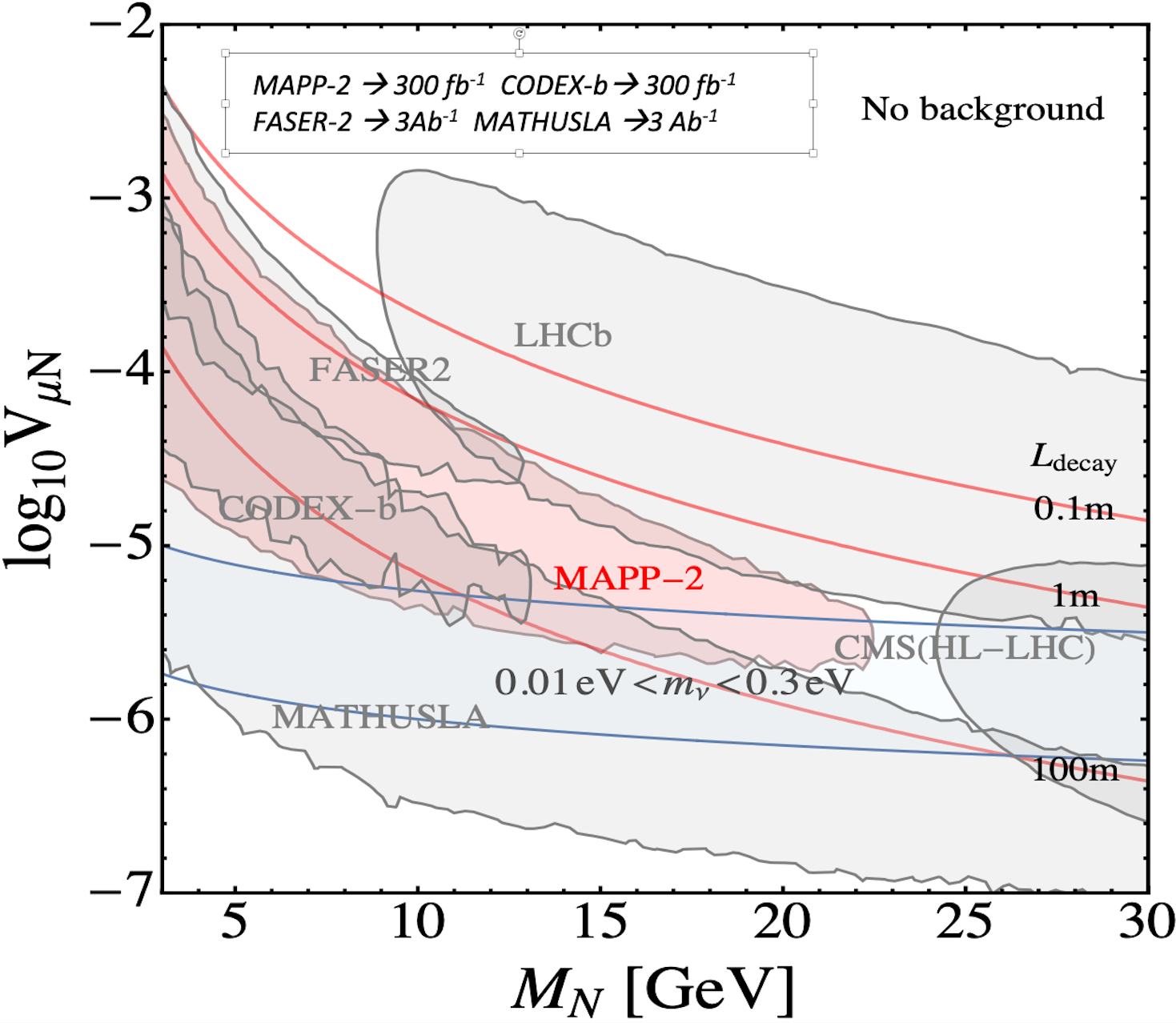}
\caption{95\% C.L. exclusion bounds estimated for  MAPP-2, and other current and proposed LHC experiments, for pair-produced HNLs in the minimal $Z' (B - L)$ model at a centre-of-mass energy of 14~TeV (Obtained via private communications with F.~Deppisch  \cite{DEPPISCH}. The $U(1)_{B-L}$ gauge coupling and the $Z'$ mass are chosen as $g_1' = 10^{-3}$ and $m_{Z'} = 3.33~m_N$, respectively. The red lines correspond to the proper decay length of the HNL, while the blue band denotes a preferred region of parameter space connected to a canonical see-saw mechanism for neutrino mass acquisition.}\label{fig:Deppisch}
\end{figure}

\noindent
MAPP-2's sensitivity to heavy neutral leptons was also studied. HNLs arise in many SM extensions containing  massive neutrinos. The gauged $B - L$ model described in Ref.~\cite{DEPPISCH} is an example of one such scenario. In this case the model requires a new Abelian $U(1)_{B - L}$ gauge field,  called $B'_{\mu}$, to be introduced into the SM along with three right-handed (RH) Majorana neutrinos $N_{i}$ and a SM singlet scalar field $\chi$. The new scalar and neutrino fields have $B - L$  charges $+2$ and $-1$,  respectively. The new physics particles contained in this model are a $Z'$ gauge boson with coupling $g'_{1}$  and three new neutrinos.  The new neutrinos, which may be heavy and long-lived, can be  produced from a $Z'$ boson,  via $Z' \rightarrow N_{i}N_{i}$, acting as a portal mediator for the particle.  The new RH Majorana neutrinos can decay to, e.g., $N \rightarrow \mu^{\pm}q\overline{q}$ and $N \rightarrow \mu^{+}\mu^{-}\nu_{\mu}$. Such displaced vertices can be detected by MAPP-2.  In Figure~\ref{fig:Deppisch}, we see again the competitive and complementary reach of the MAPP-2 detector as compared with other detectors such as  FASER2 \cite{FASER2} , CMS \cite{DEPPISCH}, CODEX-b \cite{CODEX-b-limits},  LHCb \cite{DEPPISCH} and MATHUSLA \cite{MATHUSLA}. 

In other models, sterile neutrinos could be produced in leptonic and semi-leptonic decays of charmed and bottomed mesons, decaying to leptons via neutral and charged weak currents, detectable in MAPP-2~\cite{DeVries:2020jbs}. R-parity violating supersymmetry also predicts LLPs, such as light long-lived neutralinos decaying to charged particles. Benchmark scenarios related to either charm or bottom mesons decaying into neutralinos have been considered, in similar fashion as in sterile neutrinos, showing that MAPP can cover a range of  lifetimes~\cite{Dreiner:2020qbi}.

\begin{figure}[hbt]
\centering\includegraphics[width=0.8\linewidth]{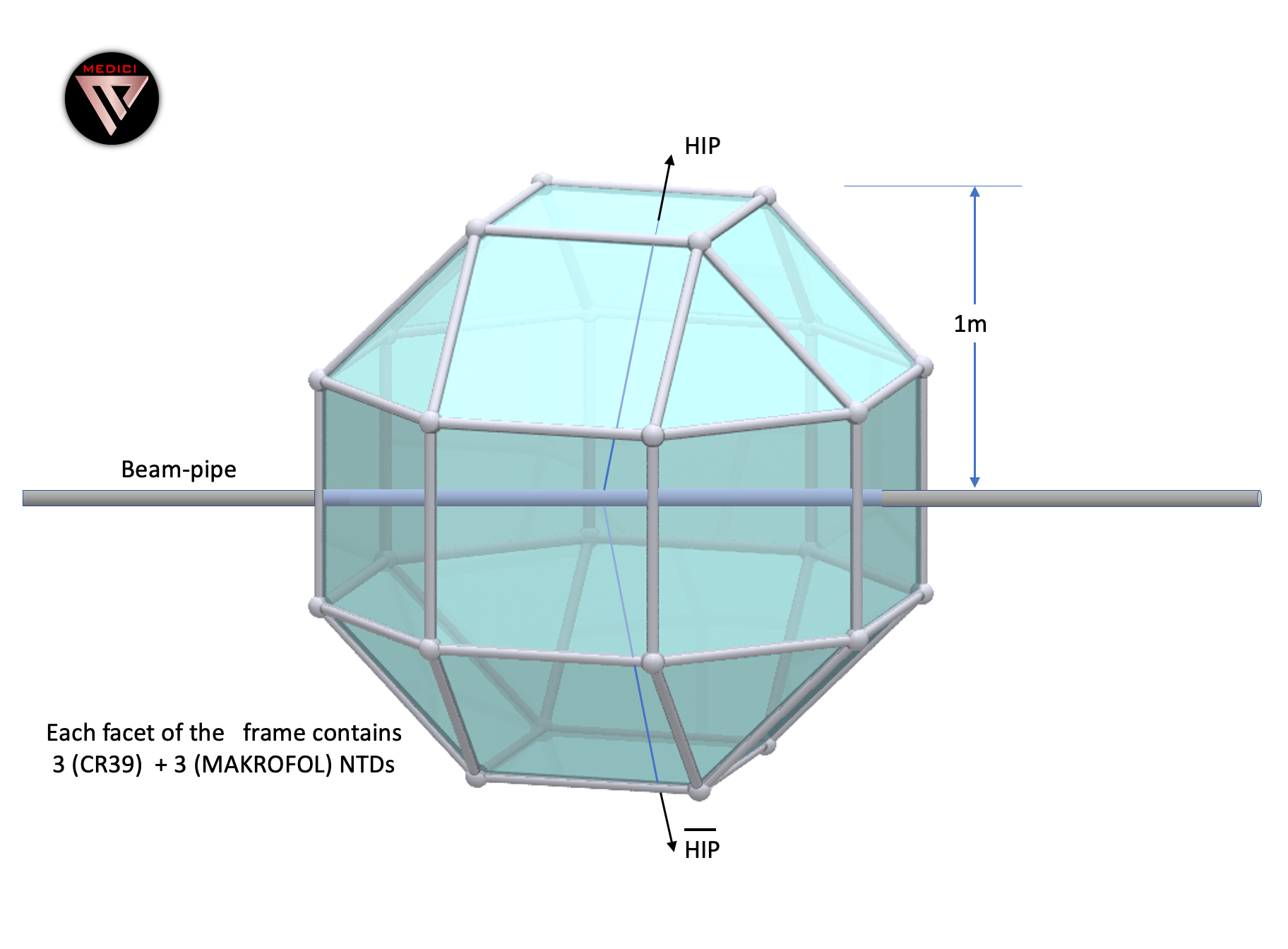}
\caption{The  MEDICI HIP detector envisaged for the FCC at 100~TeV.}
\label{fig:MEDICI-HIP-Detector}
\end{figure}

\section{Beyond the LHC -- the FCC at 100~TeV}
\noindent
 An important step for the future development of  high-energy physics is  a ${\cal O}(100$~TeV) $pp$ collider that will   follow  the completion of the LHC and HL-LHC physics programmes. In particular, CERN is studying a Future Circular Collider facility that will deliver 100~TeV $pp$ collisions using a 100~km tunnel. This facility is also designed   to deliver  e$^{+}$e$^{-}$ and $ep$ collisions, as well as  a program with heavy-ion beams and with the injector complex. We report here some preliminary plans for the deployment of  the MEDICI (Monopole and Exotics Detector Infrastructure for Colliding Ions)  for a Dedicated Search Detector Facility at the FCC-hh that we envisage will carry on the program of the MoEDAL-MAPP experiment at an $E_\text{CM}$   of 100~TeV.  The envisaged  MEDICI detectors for  HIPs and the LLPs  is sketched in Figure~\ref{fig:MEDICI-HIP-Detector}. We sketch some aspects of MEDICI's  possible physics reach next.
 
 \begin{figure}[hbt]
\centering\includegraphics[width=0.8\linewidth]{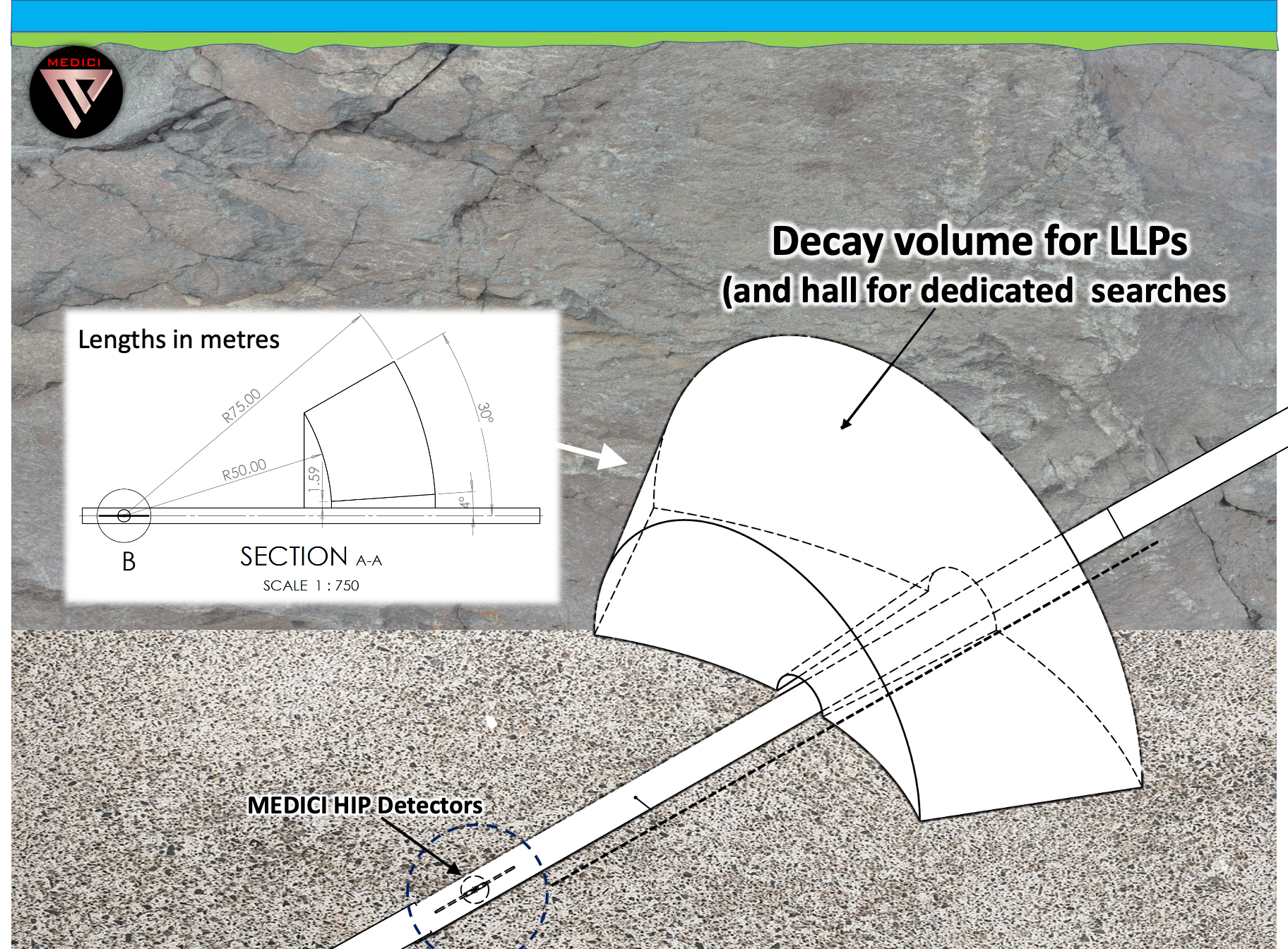}
\caption{The  MEDICI experimental cavern at the FCC-hh,  primarily  designed as a decays zone for an LLP  detector.  It is envisaged that other direct search experiments would use this cavern.}
\label{fig:Medici-LLP}
\end{figure}
 
\subsection{The Physics Reach of the MEDICI HIP Detector}
\noindent
The MEDICI HIP detector  takes the form of a polyhedral ``ball" with radius 1~m. The passive detector technology utilized is the same as that used in MoEDAL.  The big difference being that the detector is placed directly around the IP with no other intervening detector material. Assuming Drell-Yan production on monopole-pairs and using the analysis procedures, parameters and calibration  used by MoEDAL for the LHC the mass limits that we can reach for 3~ab$^{-1}$ at the 100~TeV FCC-hh machine are shown in Table~\ref{tab:monopole-lims}.
\begin{table*}[!hbt]\centering
\caption{\label{tab:monopole-lims} Mass limits for DY production on magnetic monopole pairs  at the 100~TeV FCC-hh machine. }
\vspace{0.5cm}
\begin{tabular}{|c|cccccc|} \hline
       & \multicolumn{6}{c|}{Magnetic charge ($\gd$)} \\
       & 1      &   2   &     3  &   4  & 5 & 6    \\ \hline
\colrule
Spin   & \multicolumn{6}{c|}{95\% CL mass limits [GeV/$c^{2}$]} \\
\colrule
0      & 14.9    & 17.0   &   18.2   &  19.1 & 19.8 & 20.3          \\  
1/2    & 20.0   & 22.4   &   23.7   &  24.7 & 25.5 & 26.1           \\
1     &   20.5  &  22.7   &   23.9   &  24.8 & 25.5  & 26.1          \\   \hline
\end{tabular}
\end{table*}

\noindent
As can be seen from Table~\ref{tab:monopole-lims} at the 100~TeV FCC-hh machine one can push the search for MMs up
to around 25~Tev/$c^{2}$, depending on the charge and spin of the magnetic monopole.

\subsection{Searching for LLPs with MEDICI}
\noindent
In order to assess the physics reach in the search for LLPs at the 100~TeV FCC-hh machine \cite{Mangano} we introduce the MEDICI underground decay volume shown in Figure~\ref{fig:MEDICI-HIP-Detector} some 50~m for an interaction point of the FCC ring.  The decay volume ``experimental hall'' is engineered in the shape of a half volume element in spherical coordinates in order to minimize the amount of material that has to be removed. As can be seen from the figure the hall is 37.5~m high at its highest point and 50~m from the MEDICI IP on FCC ring. In order to provide a practical scale for the  size of the decay volume we require that it provides a slightly better sensitivity at the longest lifetimes for the decay of dark Higgs bosons \cite{Feng-2018,Gligorov-2018}  benchmark,  than that of the MATHUSLA detector  at the HL-LHC at the longest lifetimes. 

\begin{figure}[hbt]
\centering\includegraphics[width=0.7\linewidth]{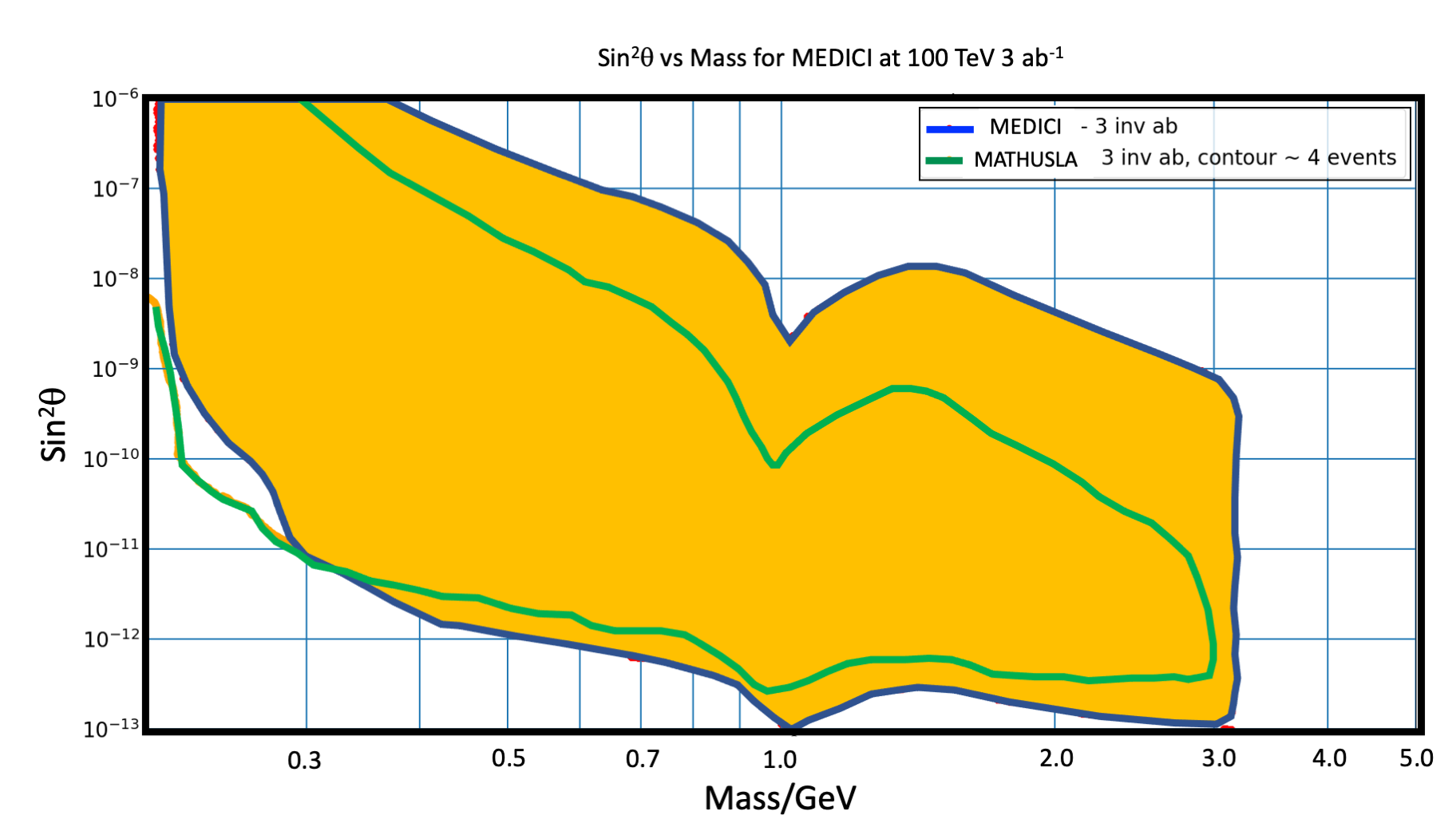}
\caption{The sensitivity contours using the well  studied  benchmark involving the decay of dark Higgs bosons \cite{Feng-2018,Gligorov-2018} that was described previously obtained the MEDICI detector operating at and $E_\text{CM}$ of 100~TeV at the FCC-hh collider.}
\label{fig:Medici-LLP-scalar-portal}
\end{figure}

\noindent
The sensitivity contour obtained for 3~ab$^{-1}$ of data are shown in Figure~\ref{fig:Medici-LLP-scalar-portal} using the dark Higgs boson
benchmark considered previously. We use as a benchmark the sensitivity contour obtained for the MATHUSLA detector operating at the  HL-LHC.
As can be seen from the figure the exclusion contour for the MATHUSLA detector is completely enclosed in that of the MEDICI LLP detector. The MEDICI 
detector will be at least 100~m underground which severely reduces the background levels from cosmic rays that would be experienced by the 
MATHUSLA detector deployed on the surface. We estimate that the cost of the MEDICI LLP detector and its experimental hall is of the order of \$10M.

\section{Conclusion and Summary} 
\noindent
The MoEDAL-MAPP project, that is planned in three phases  extending  over  the lifetime of the LHC machine (2010 $\rightarrow$ 2037). Phase-0 of the project began in 2010 with the deployment of the MoEDAL detector prototype. During  LHC's Run-1\&2 (2011-2018) MoEDAL,  the LHC's first dedicated search experiment, took data with 100\% efficiency at IP8. Designed to search for highly ionizing particle avatars  of New Physics, MoEDAL's physics program  defines over 34 scenarios yielding potentially revolutionary insights into foundational questions, involving: magnetic charge; $4^{\rm th}$ generation; heavy neutrinos; extra dimensions; new symmetries such as SUSY; and the nature of dark matter.

\noindent
Since its approval in 2010, the MoEDAL experiment has placed the world's best limits on the existence of singly and multiply charged magnetic monopoles and highly electrically charged objects.  It has also carried out the first ever searches for: (a) Spin-1 MMs; (b) dyons, particles with electric and magnetic charge; and, (c) MMs produced in heavy-ion collisions via the Schwinger mechanism (published in \emph{Nature}~\cite{MoEDAL-2022}). Additionally, it has presented the LHC's only search for monopole production via photon fusion. The continuing analysis of its Run-2 data promises other pioneering results.

\noindent
In 2021, CERN's Research Board unanimously approved the reinstallation of an updated MoEDAL detector at IP8 on the LHC ring~\cite{moedal-run3-TP}; and, the installation of the major MoEDAL detector upgrade MAPP (MoEDAL Apparatus for Penetrating Particles)  in the UA83 tunnel~\cite{MAPP-TP}. This marks Phase-1 of the MoEDAL-MAPP project. At the time of writing both detectors are being installed for data taking during LHC's Run-2 beginning in 2022 (MoEDAL) and in the spring of 2023 (MAPP). 

\noindent
The MoEDAL-MAPP experiment now has a much-enhanced physics reach with sensitivity not only to highly ionizing particles, but also to feebly interacting particles, such as milli-charged particles, and very long-lived particles. The MoEDAL-MAPP experiment provides a complementary expansion of the LHC's discovery horizon by providing sensitivity to scores of new physics scenarios for which the main LHC experiments are not optimized, that involve HIPS, FIPs and LLPs.
 
\noindent
 We are redeploying MoEDAL for data  taking at Run-3 with a estimated factor of five greater instantaneous  luminosity, greater efficiency for  the NTD detectors and a lower detection threshold falling  from    $\sim$ 50$e$  to $\sim$ 5$e$ by utilizing MoEDAL's CR39    detectors that are now  fully calibrated.  The luminosity available during Run-3 and HL-LHC allow us to push the quest for  MMs and HECOs to higher mass, higher charge and lower cross-section. Importantly, we can now push the search for electrically charged particles that have charge less than 10$e$. 
 
 \noindent
 As we have seen from the examples discussed above such particles stem from a number of  BSM scenarios including for example: supersymmetry, neutrino mass models,  L-R symmetric models, etc. Typically, MoEDAL has a complementary and  competitive sensitivity for higher mass long-lived states. MoEDAL's totally different systematics together with is lack of SM backgrounds, permanent  physical record of the signal,  and its  trigger-less operation would make it an invaluable potential contributor to the confirmation of any discovery made in this arena. 
 
 \noindent
 The MAPP upgrade to the MoEDAL detector now  approved by CERN's Research Board is currently being installed in the UA83 gallery  some 100~m from IP8. The MoEDAL-MAPP experiment now has a much-enhanced physics reach with  a challenging  sensitivity not only to Highly Ionizing Particles, but also to Feebly Interacting Particles, such as Milli-Charged Particles, and very Long-Lived Particles. The MoEDAL-MAPP experiment provides a complementary expansion of the LHC's discovery horizon by providing sensitivity to scores of new physics scenarios  for which the main LHC experiments are not optimized, for example from dark sector models that involve HIPS, FIPs and LLPs. The  cost, ignoring person-power costs,  to design, construct and instal the MAPP-1 detector is less than 1M CHF.
 
 \noindent
 The MoEDAL-MAPP collaboration is already planning Phase-2 of their project to be installed in LS3 in 2025-2026 for first data taking in 2027-2028. At this stage MAPP-2 will be installed in the UGC1 gallery adjacent to IP8, joining the MoEDAL and MAPP detectors as part of the MoEDAL-MAPP facility. Initial physics studies of MAPP-2  performance indicates that it make major contributions to the search for LLPs in a way that is complementary and competitive to both the main LHC detectors and the high cost dedicated detector projects (~\$100M) such as MATHUSLA and SHiP. A civil engineering study of the UGC1 gallery was commissioned by MoEDAL-MAPP in 2021 with a view to bringing it up to standard as an LHC experimental area. This study together with preliminary design estimates gives us an initial cost estimate for the MAPP-2 detector of 5-6M CHF.
 
\noindent 
 MoEDAL-MAPP's forward planning group is already outlining a cost-effective dedicated detector  facility for the contemplated 100~TeV FCC-hh machine. We envisage here  a dedicated open interaction point. The interaction point will also be housed in a small experimental hall. Downstream of this a larger experimental hall, although smaller than those on the LHC ring,  acts as a decay zone for LLPs. We expect that  MEDICI facility will also service other dedicated search detectors, yet to be proposed.  Initial physics studies have shown that an NTD ``Ball'' detector could quest for monopoles  as massive as 20-30~TeV/$c^{2}$.  Also, with the decay zone described above we would, for the baseline ``dark scalar decay to muon-pair '' scenario considered above,  have essentially the same sensitivity for LLPs as MAPP-2 and MATHUSLA combined.


\section{Acknowledgments}
\noindent
We thank CERN for the  LHC's successful operation, as well as the support staff from our institutions without whom MoEDAL could not be operated. We acknowledge the invaluable assistance of  particular members of the LHCb Collaboration: G. Wilkinson, R. Lindner, E.  Thomas and G. Corti. In addition we would like to recognize the valuable input from W-Y Song and W. Taylor of York University on HECO production processes.   Computing support was provided by the GridPP Collaboration, in particular by the Queen Mary University of London and Liverpool grid sites. This work was supported by grant PP00P2\_150583 of the Swiss NSF; by the UK Science and Technology Facilities Council, via the grants, ST/L000326/1, ST/L00044X/1, ST/N00101X/1, ST/P000258/1 and ST/T000759/1; by the Generalitat Valenciana via a special grant for MoEDAL and via the projects PROMETEO/2019/087, CIPROM/2021/054 and CIPROM/2021/073; by MCIU / AEI / FEDER, UE via the grants PGC2018-094856-B-I00, PID2020-113334GB-I00 and PID2021-122134NB-C21; by the Physics Department of King's College London; by  NSERC via a project grant; by the V-P Research of the University of Alberta (UofA); by the Provost of the UofA); by IFA (Romania); by the INFN (Italy); by the Estonian Research Council via a Mobilitas Plus grant MOBTT5;  by a National Science Foundation grant (US) to the University of Alabama MoEDAL group; and, by the National Science Centre, Poland, under research grant 2017/26/E/ST2/00135 and the Norwegian Financial Mechanism for years 2014-2021, grant DEC-2019/34/H/ST2/00707.


\end{document}